\documentclass[12pt,a4paper]{article}
\usepackage[truedimen,margin=30mm]{geometry}

\normalsize

\usepackage{mathrsfs}
\usepackage{amssymb}
\usepackage{amsmath}
\usepackage{ascmac}
\usepackage{amsthm}
\usepackage[dvips]{graphicx}
\usepackage{natbib}
\usepackage{setspace}
\usepackage{times}
\usepackage{color}
\usepackage{url}
\usepackage{pdflscape}   
\usepackage{comment}   
\usepackage{bm}

\newcommand{\indep}{\mathop{\perp\!\!\!\perp}}

\newcommand{\bld}{\boldsymbol}

\usepackage{titlesec}
\titleformat*{\section}{\large\bfseries}
\titleformat*{\subsection}{\it}

\allowdisplaybreaks[4]

\newtheorem{thm}{Theorem}
\newtheorem{lem}{Lemma}

\newtheorem{prp}{Proposition}

\newtheorem{algo}{Algorithm}

\newtheorem{rem}{Remark}

\newtheorem{alem}{Lemma A.\!}
\newtheorem{aalgo}{Algorithm A.\!}

\numberwithin{equation}{section}

\def\ep{{\varepsilon}}

\def\bX{\bm{X}}

\def\balp{\bm{\alpha}}
\def\bbe{\bm{\beta}}

\title{{\bf Bayesian Doubly Robust Causal Inference via Posterior Coupling}\footnote{\today}}

\date{}

\begin{document}

\maketitle
\doublespacing

\vspace{-1.5cm}
\begin{center}
{\large Shunichiro Orihara$^1$, Tomotaka Momozaki$^2$ and Shonosuke Sugasawa$^3$}

\medskip

\medskip
\noindent
$^1$Department of Health Data Science, Tokyo Medical University\\
$^2$Department of Information Sciences, Tokyo University of Science\\
$^3$Faculty of Economics, Keio University
\end{center}

\vspace{0.5cm}
\begin{center}
{\bf \large Abstract}
\end{center}
Bayesian doubly robust (DR) causal inference faces a fundamental dilemma:\ joint modeling of outcome and propensity score suffers from the feedback problem where outcome information contaminates propensity score estimation, while two-step inference sacrifices valid posterior distributions for computational convenience. 
We resolve this dilemma through posterior coupling via entropic tilting. 
Our framework constructs independent posteriors for propensity score and outcome  models, then couples them using entropic tilting to enforce the DR moment condition. 
This yields the first fully Bayesian DR estimator with an explicit posterior distribution. 
Theoretically, we establish three key properties:\ (i) when the outcome model is correctly specified, the tilted posterior coincides with the original; (ii) under propensity score model correctness, the posterior mean remains consistent despite outcome model misspecification; (iii) convergence rates improve for nonparametric outcome models. 
Simulations demonstrate superior bias reduction and efficiency compared to existing methods. 
We illustrate practical advantages of the proposed method through two applications:\ sensitivity analysis for unmeasured confounding in antihypertensive treatment effects on dementia, and high-dimensional confounder selection combining shrinkage priors with modified moment conditions for right heart catheterization mortality. 
We provide an R package implementing the proposed method. 
\vspace{-0cm}


\section{Introduction}










In observational studies, estimating causal effects while adjusting for confounders is a fundamental task. The propensity score plays a central role in this context \citep{rosenbaum1983central}, particularly for estimating the average treatment effect (ATE). 
Among propensity score based methods, the inverse probability weighting (IPW) estimator and its extension, known as the augmented IPW (AIPW) estimator, are widely adopted. 
In particular, AIPW estimator incorporates information from an outcome model \citep{tsiatis2006semiparametric} and is known as a doubly robust (DR) estimator.
The DR estimator possesses a key property known as ``double robustness", meaning that it remains consistent if either (1) the propensity score model or (2) the outcome model is correctly specified. This property makes the DR estimator particularly appealing in practice.

In Bayesian contexts, causal inference has gained increasing attention in recent years \citep{daniels2023bayesian}. A commonly applied approach is based on the G-formula, which is relatively easy to interpret within the Bayesian framework, as it relies on likelihoods for the outcome and confounders. For this reason, G-formula based methods typically do not require the use of propensity score information. As discussed in \cite{saarela2016bayesian}, incorporating propensity score information can improve the robustness of estimators, which is the same motivation underlying the DR estimator \citep{zhang2009extensions}. This property provides a compelling reason to consider the use of propensity scores in Bayesian causal inference. In Bayesian contexts, the propensity score is typically included within the likelihood for the outcome model. However, this inclusion gives rise to a well-known issue referred to as the feedback problem \citep{li2023bayesian, stephens2023causal}, whereby the estimated propensity score may fail to adequately adjust for confounding \citep{saarela2016bayesian}. 
As a result, it is often difficult to avoid this issue when incorporating propensity score information into model construction.

In previous studies, Bayesian DR estimation methods have been proposed while avoiding the cutting feedback problem. \cite{saarela2016bayesian} proposed a Bayesian DR estimator by introducing a weighting loss function for the estimator, combined with the Bayesian bootstrap. This approach can be interpreted as defining a loss function for a pseudo-population in which confounding effects are removed \citep{He2020}. While this method represents an important contribution from the Bayesian perspective, it does not yield an explicit posterior distribution. \cite{antonelli2022causal} also proposed another Bayesian DR method that incorporate propensity score information in a thoughtful way; however, this approach similarly lack tractable expressions for the posterior distribution.

In this manuscript, we propose a novel framework for Bayesian DR inference that provides an explicit posterior distribution. 
Our key innovation is to couple separate (independent) posterior distributions for the outcome (i.e., excluding propensity score information) and propensity score models. 
In our proposed method, propensity score information is incorporated into the outcome model using entropic tilting (ET) \citep{jaynes1957information,tallman2022entropic} based on a moment condition that corresponds to the augmentation term in the DR estimator \citep{Bang2005doubly}, ensuring double robustness. 
While \cite{yiu2020inference} and \cite{luo2023semiparametric} employ a similar tilting idea, our approach differs in that it yields an explicit posterior distribution with a relatively simple moment condition and procedure.
The recent work by \cite{breunig2025double} also develops a nonparametric Bayesian DR method, but requires auxiliary data. 

The explicit posterior formulation of our method facilitates natural extensions to challenging problems in causal inference. 
We demonstrate how the framework readily accommodates sensitivity analysis for unmeasured confounding through importance weighting, allowing practitioners to assess the robustness of causal conclusions under violations of the ignorability assumption. 
Additionally, we show how the method can be extended to high-dimensional settings through a novel combination of shrinkage priors and modified moment conditions that focus on selected confounders, addressing regularization-induced confounding by reviving previously overlooked weak confounders through entropic tilting. 
These extensions illustrate the versatility of our framework beyond standard DR estimation. 

The remainder of this manuscript is organized as follows. 
Section \ref{sec:background} briefly introduces the notation for causal inference and reviews standard DR estimators from non-Bayesian perspectives. 
Section \ref{sec:proposed} presents the proposed Bayesian DR estimator via posterior coupling and provides efficient computational algorithms. 
Section \ref{sec:theorems} establishes theoretical robustness properties of the proposed method. 
Section \ref{sec:sim} conducts simulation experiments to evaluate the performance compared with existing methods. 
Section \ref{sec:sensitivity} demonstrates the application to dementia prevention with sensitivity analysis for unmeasured confounding. 
Section \ref{sec:selection} presents an application to mortality prediction with confounder selection in high-dimensional settings. 
Section \ref{sec:discussion} provides concluding remarks. 
All technical proofs and additional simulation results are provided in the Supplementary Material.



\section{Background} \label{sec:background}

\subsection{Notations and two causal effect estimators}\label{sec2.1}

Let $(Y_i, A_i, \bX_i)$ be a triplet of the observed data for $i=1,\ldots,n$, where $Y_i$ is an outcome, $A_i$ is a binary treatment indicator, and $\bX_i$ is a vector of covariates. Let $(Y_{1i}, Y_{0i})$ be potential outcomes under treated $(Y_{1i})$ and control $(Y_{0i})$, and $Y_i$ can be expressed as $Y_i=A_iY_{1i} + (1-A_i)Y_{0i}$. Note that the outcome variable is not limited to being continuous; it can also be categorical or ordinal in the following discussions. The estimand of interest in this work is the average treatment effect (ATE), defined as $\tau:=E[Y_{1i}-Y_{0i}]$. For notational simplicity, the subscript $i$ may be omitted when it is clear from the context.

To estimate the ATE, strong ignorability treatment assignment is commonly assumed:\ $(Y_{1}, Y_{0}) \indep A \mid \bX$ \citep{rosenbaum1983central}. Here, the covariates $\bX$ are sometimes referred to as confounders. Under this assumption, several methods for estimating the ATE can be considered. One basic method is regression adjustment. Using a regression model $f(y \mid A_i = a, \bX_i)$, the ATE is given by the following expectation:
\begin{align}
\label{reg1}
\frac{1}{n}\sum_{i=1}^{n}\left\{{\rm E}\left[Y\mid A_{i}=1,\bX_{i}\right]-{\rm E}\left[Y\mid A_{i}=0,\bX_{i}\right]\right\}=:\frac{1}{n}\sum_{i=1}^{n}\left\{m_{1}(\bX_{i})-m_{0}(\bX_{i})\right\},
\end{align}
where $m_a(\bX_i)={\rm E}\left[Y\mid A_{i}=a,\bX_{i}\right]$ is the outcome model. 
Another common approach is to use the propensity score, defined as $e(\bX_i) = {\rm Pr}(A = 1 \mid \bX_i)$ \citep{rosenbaum1983central}. In particular, the following inverse probability weighting (IPW) estimator is often considered:
\begin{align}
\label{ipw1}
\frac{1}{n}\sum_{i=1}^{n}\left\{\frac{A_{i}Y_{i}}{e(\bX_{i})}-\frac{(1-A_{i})Y_{i}}{1-e(\bX_{i})}\right\}.
\end{align}

\subsection{Doubly robust estimator}\label{sec2.2}
In causal inference contexts, the doubly robust (DR) estimator \citep{tsiatis2006semiparametric}, which combines regression adjustment (\ref{reg1}) and the IPW estimator (\ref{ipw1}), is widely considered. An attractive feature of the DR estimator is its consistency for the ATE if either the outcome model or the propensity score model is correctly specified---but not necessarily both. Specifically:
\begin{align}
\label{dr1}
\frac{1}{n}\sum_{i=1}^{n}\left\{m_{1}(\bX_{i})-m_{0}(\bX_{i})+\frac{A_{i}}{e(\bX_{i})}\left(Y_{i}-m_{1}(\bX_{i})\right)-\frac{1-A_{i}}{1-e(\bX_{i})}\left(Y_{i}-m_{0}(\bX_{i})\right)\right\}&\nonumber\\
&\hspace{-13cm}=\frac{1}{n}\sum_{i=1}^{n}\left\{m_{1}(\bX_{i})-m_{0}(\bX_{i})+\frac{A_{i}-e(\bX_{i})}{e(\bX_{i})(1-e(\bX_{i}))}\left(Y_{i}-m_{A_{i}}(\bX_{i})\right)\right\}.
\end{align}

\cite{Bang2005doubly} propose another DR estimator that estimates the regression model $m_{A_{i}}(\bX_{i}; \bbe)$ with parameter $\bbe$ by setting the third term of (\ref{dr1}) to zero. Because the third term is set to zero, the regression model (\ref{reg1}) using the estimated model $m_{A_{i}}(\bX_{i}; \bbe)$ achieves double robustness. This concept is a central idea in the following discussion of our manuscript.

\subsection{Bayesian approaches to doubly robust estimation}\label{sec2.3}
In spite of the popularity of estimators using the propensity score, such as the IPW estimator and the DR estimator, Bayesian interpretation of these estimators are challenging due to two conflicting approaches, ``joint modeling" and ``cutting feedback".

In terms of Bayesian modeling, constructing the joint distribution of the outcome model and the propensity score model, refereed to as ``joint modeling", would be a natural approach. 
Let $e(\bX_i; \balp)$ be a propensity score model with parameter $\balp$ and $f(Y_i\mid A_i, \bX_i; e(\bX_i;\balp), \bbe)$ be a outcome model dependent on $e(\bX_i;\balp)$ and parameter $\bbe$. 
Assuming the prior independence of $\balp$ and $\bbe$, the joint posterior distribution of $Y_i$ and $A_i$ can be obtained as 
\begin{equation}\label{eq:joint-model}
\pi(\balp)\pi(\bbe)\prod_{i=1}^n f(Y_i\mid A_i, \bX_i; e(\bX_i;\balp), \bbe) e(\bX_i; \balp)^{A_i}\{1-e(\bX_i; \balp)\}^{1-A_i},
\end{equation}
where $\pi(\balp)$ and $\pi(\bbe)$ are prior distributions of $\balp$ and $\bbe$, respectively. 
A notable property of the above posterior is that the (marginal) posterior distribution of $\balp$ includes information of the outcome $Y_i$ through the propensity score in the outcome model $f(Y_i\mid A_i, \bX_i; e(\bX_i;\balp), \bbe)$, even through $\balp$ and $\bbe$ are independent in the prior distribution. 
This is not consistent with the philosophy of constructing propensity scores and may deteriorate its balancing property \citep{saarela2016bayesian}.

On the other hand, it would be natural to estimate the propensity score first, and then estimate causal effects given the estimated propensity score \citep{imbens2015causal}.
This approach is called ``cutting feedback" \citep{li2023bayesian, stephens2023causal} since the posterior of $\balp$ is constructed by explicitly removing feedback from the outcome model in the joint posterior (\ref{eq:joint-model}). 
While such two-step inference would be natural in a non-Bayesian approaches, it may not give a valid posterior distribution.

To address this, \cite{saarela2016bayesian} considers modeling the joint distribution for the outcome and propensity score models, ensuring that the outcome model does not include information from the propensity score model. We follow this idea in the construction of the general posterior introduced in Section \ref{sec:proposed}. 
The critical difference between our proposed method and previous works on Bayesian DR estimation \citep{saarela2016bayesian, yiu2020inference,antonelli2022causal, luo2023semiparametric, breunig2025double} is that our approach constitutes a fully Bayesian estimation. Specifically, our method involves constructing the (general) posterior distribution of $\balp$ and $\bbe$, separately, and then modifying it through constraint to ensure double robustness.

\section{Bayesian doubly robust inference} \label{sec:proposed}

\subsection{Separate construction of posterior distributions}\label{sec3.1}
In this manuscript, we consider a Bayesian approach using the general posterior distribution \citep[e.g.][]{yin2009bayesian, bissiri2016general}. Specifically, we consider the following pseudo-likelihood functions for the outcome regression model and the propensity score model, respectively:
\begin{align}
\ell(\bbe)&= \exp\left\{-nf_n(\bbe)\right\}, \ \ \ \ f_n(\bbe)=\frac1n\sum_{i=1}^n f(Y_{i} \mid  A_{i},\bX_{i};\bbe),
\label{gl1}\\
\ell(\balp)&= \exp\left\{- nf_n(\balp)\right\}, \ \ \ \ 
f_n(\balp)=\frac1n\sum_{i=1}^nf(A_{i} \mid \bX_{i};\balp)\label{gl2}
\end{align}
where $\balp\in\Theta_{\balp}$ and $\bbe\in\Theta_{\bbe}$.
Note that when $nf_n(\balp)$ and $nf_n(\bbe)$ are negative log-likelihood of a unit sample, both (\ref{gl1}) and (\ref{gl2}) reduce to the standard likelihood functions.  
The outcome model (\ref{gl1}) could be constructed ``model-free" squared loss function, $f(Y_{i}\mid A_{i},\bX_{i};\bbe)=\omega \left(Y_{i}-m_{A_{i}}(\bX_{i};\bbe)\right)^2$, 
where $\omega$ is a learning rate \citep[e.g.][]{bissiri2016general,wu2023comparison}. 
For the propensity score model (\ref{gl2}), a standard logistic regression model for the propensity score is equivalent to specifying $f(A_{i} \mid \bX_{i};\balp)=A_i\log\{e(\bX_i;\balp)\}+(1-A_i) \log\{1-e(\bX_i;\balp)\}$ with propensity score $e(\bX_i;\balp)$.
It can also be derived from the covariate balancing propensity score conditions \citep{imai2014covariate, orihara2024general}. 
In the following discussions, we assume that global maximizers of (\ref{gl1}) and (\ref{gl2}) exist, denoted by $\balp^{*}$ and $\bbe^{*}$, respectively. 

Given prior distributions on $\balp$ and $\bbe$, the joint general posterior distribution of $(\balp, \bbe)$ given the observed data $D := \{(Y_i, A_i, \bX_i), i=1,\ldots,n\}$ 
is expressed as
\begin{equation}\label{eq:pos}
p_{n}(\balp, \bbe\mid D)
=
\frac{p(\balp)p(\bbe)\exp\left\{-nf_{n}(\balp)\right\}\exp\left\{-nf_{n}(\bbe)\right\}
}
{\iint p(\balp)p(\bbe)\exp\left\{-nf_{n}(\balp)\right\}\exp\left\{-nf_{n}(\bbe)\right\}d\balp d\bbe
},
\end{equation}
where $p(\balp)$ and $p(\bbe)$ are prior distributions of $\balp$ and $\bbe$, respectively. 
Due to the form of the general posterior distribution, the joint posterior (\ref{eq:pos}) can be decomposed as $p_{n}(\balp, \bbe\mid D)=p_{n}(\balp\mid D)p_{n}(\bbe\mid D)$ \citep{gelman1995bayesian}, where $p_{n}(\balp, \mid D)\propto p(\balp)\exp\left\{-nf_{n}(\balp)\right\}$ and $p_{n}(\bbe\mid D)\propto p(\bbe)\exp\left\{-nf_{n}(\bbe)\right\}$.
A notable feature of the posterior (\ref{eq:pos}) is that the posterior of $(\balp,\bbe)$ are separately constructed unlike the existing Bayesian approaches that includes a propensity score model in the outcome model \citep[e.g.][]{saarela2016bayesian}, leading to the joint posterior in which $\balp$ and $\bbe$ are correlated. 
While inclusion of a propensity score model in the outcome model complicates the posterior computation of the joint posterior, the posterior (\ref{eq:pos}) can be easily constructed since propensity and outcome models are separately estimated.


For the general posteriors, the following posterior concentration property holds.

\begin{prp} \citep{miller2021asymptotic} \ Under some regularity conditions, 
$$
\int_{\balp \in A_{\varepsilon}^{\balp}} p_n(\balp\mid D)d\balp \to 1\ \ \ and\ \ \ \int_{\bbe \in A_{\varepsilon}^{\bbe}}p_n(\bbe\mid D)d\bbe\to 1,
$$
where $A_{\varepsilon}^{\balp}=\left\{\balp\in\Theta_{\balp}: f(\balp)<f(\balp^{*})+\epsilon\right\}$ and $A_{\varepsilon}^{\bbe}=\left\{\bbe\in\Theta_{\bbe}: f(\bbe)<f(\bbe^{*})+\epsilon\right\}$ for all $\varepsilon>0$, and $f_{n}(\balp)\to f(\balp)$ and $f_{n}(\bbe)\to f(\bbe)$ for all $\balp\in\Theta_{\balp}$ and $\bbe\in\Theta_{\bbe}$, respectively. 
\end{prp}

\noindent
From the proposition, when the outcome model is correctly specified, denoted as $\bbe^{*}=\bbe^{0}$ (i.e., ${\rm E}\left[Y\mid A_{i},\bX_{i}\right]=m_{A_{i}}(\bX_{i};\bbe^{0})$), the regression-based estimator is valid for estimating the ATE. However, if the model is misspecified, it is no longer valid. Our objective is to construct a DR-like Bayesian estimator that leverages the propensity score information, even when the outcome model is misspecified.

\subsection{Combining propensity score and outcome models via posterior coupling}\label{sec3.2}

To construct DR posterior, we couple information of two posteriors based on outcome and propensity score models. 
Specifically, we employ the entropic tilting \citep{jaynes1957information,tallman2022entropic}, to obtain constraint posterior distribution under pre-specified moment conditions. 
In particular, we employ the constraint $B_n(\balp,\bbe)=0$, where 
\begin{equation}\label{eq:balancing}
B_n(\balp,\bbe)\equiv \frac1n\sum_{i=1}^n \frac{A_{i}-e(\bX_{i};\balp)}{e(\bX_{i};\balp)(1-e(\bX_{i};\balp))}\left(Y_{i}-m_{A_{i}}(\bX_{i};\bbe)\right).
\end{equation}
This term corresponds to the third term of the ordinary DR estimator (\ref{dr1}).
We then propose modifying the original posterior (\ref{eq:pos}) such that the posterior mean of (\ref{eq:balancing}) becomes zero. According to the entropic tilting framework \citep{jaynes1957information}, the optimal distribution that is closest to the original in terms of the Kullback-Leibler (KL) divergence is expressed as
\begin{equation}\label{eq:tilted-posterior}
\pi_{n,\lambda}(\balp,\bbe) 
=
\frac{\exp\left\{\lambda B_n(\balp,\bbe)\right\}p_{n}(\balp\mid D)p_{n}(\bbe\mid D)}
{\iint \exp\left\{\lambda B_n(\balp,\bbe)\right\}p_{n}(\balp\mid D)p_{n}(\bbe\mid D)d\balp d\bbe},
\end{equation}
where $\lambda$ is a scalar value determined as the solution of
\begin{equation}\label{eq:constraint}
{\rm E}_{\pi_{n,\lambda}(\balp,\bbe)}\left[B_{n}(\balp,\bbe)\right]=0.
\end{equation} 
Here, ${\rm E}_{\pi_{n,\lambda}(\balp,\bbe)}$ denotes the expectation with respect to the tilted posterior (\ref{eq:tilted-posterior}). Note that this condition is similar to the concept of the ``clever covariate" in non-Bayesian contexts. This type of condition allows the initial estimator to possess double robustness \citep{rose2008simple}.
The above construction grantees that the posterior mean of $B_n(\balp,\bbe)$ under the tilted posterior is zero. 
To solve the constraint (\ref{eq:constraint}), we need to evaluate the expectation in (\ref{eq:constraint}), which requires generating random samples from the tilted posterior. 
However, the main difficulty is that the tilted posterior with fixed $\lambda\neq 0$ would not be a familiar form. 
In the subsequent section, we will provide an efficient sampling algorithm for solving (\ref{eq:constraint}), by using a sequential Monte Carlo method.

Given $\lambda\neq 0$, $\balp$ and $\bbe$ are correlated in the tilted posterior (\ref{eq:tilted-posterior}) unlike the original posterior. 
In other words, information from the propensity score is incorporated into the outcome model. 
Using random samples of $(\balp, \bbe)$ generated from the tilted posterior, we generate random samples of the following ATE parameter: 
\begin{equation}\label{eq:ATE}
\frac{1}{n}\sum_{i=1}^{n}\left\{m_{1}(\bX_{i};\bbe)-m_{0}(\bX_{i};\bbe)\right\}.
\end{equation}
This is a standard G-formula based on an outcome model. 
The main difference from the existing approach is that the marginal posterior of $\bbe$ contains information regarding the propensity score through entropic tilting. 
Based on the posterior samples, we can obtain the following posterior mean of the ATE parameter (\ref{eq:ATE}):
\begin{equation}\label{eq:ATE-PM}
{\rm E}_{\pi_{n,\lambda}(\bbe)}\left[\frac{1}{n}\sum_{i=1}^{n}\left\{m_{1}(\bX_{i};\bbe)-m_{0}(\bX_{i};\bbe)\right\}\right].
\end{equation}
Moreover, using the random samples of (\ref{eq:ATE}), we can compute credible intervals for uncertainty quantification. 


\subsection{Algorithms for tilting posterior}\label{sec3.3}

We here provide detailed computation algorithms for the tilting parameter $\lambda$ and generating random samples from the tilted posterior (\ref{eq:tilted-posterior}).
Specifically, we provide two algorithms, importance sampling and sequential Monte Carlo.

The idea of importance sampling is rather straightforward.
Given the posterior sample $(\balp^{\star}, \bbe^{\star})$ from the original posterior, the sample from (\ref{eq:tilted-posterior}) can be obtained by re-weighting $(\balp^{\star}, \bbe^{\star})$ with importance weight proportional to $\exp\{\lambda B_n(\balp^{\star},\bbe^{\star})\}$.
Then, the equation (\ref{eq:balancing}) can be approximated as follows: 
\begin{equation}\label{eq:constraint-pos}
\begin{split}
&\frac{\sum_{s=1}^S \exp\{\lambda B_n(\balp^{(s)},\bbe^{(s)})\}B_n(\balp^{(s)},\bbe^{(s)})}{\sum_{s=1}^S \exp\{\lambda B_n(\balp^{(s)},\bbe^{(s)})\}}=0.\\
&\hspace{0.5cm} \Leftrightarrow \ \ 
\sum_{s=1}^S \exp\{\lambda B_n(\balp^{(s)},\bbe^{(s)})\}B_n(\balp^{(s)},\bbe^{(s)})=0,
\end{split}
\end{equation} 
where $(\balp^{(s)}, \bbe^{(s)})$ for $s=1,\ldots,S$ are random samples generated from the original posterior (\ref{eq:pos}).
Hence, we can easily apply the Newton-Raphson type algorithm to solve the equation (\ref{eq:constraint-pos}) as follows:

\begin{algo}[Tuning parameter selection via importance sampling]\label{algo:IS} \ Starting with an initial 
Starting with the initial value $\lambda_{(0)}=0$ and $t=0$, update the parameter value as 
$$
\lambda_{(t+1)} \ 
\leftarrow \ 
\lambda_{(t)}-\frac{\sum_{s=1}^S\exp(\lambda_{(t)} B_s)B_s}{\sum_{s=1}^S\exp(\lambda_{(t)} B_s) B_s^{2}}
,
$$
where $B_s=B_{n}(\balp^{(s)},\bbe^{(s)})$. 
The updating process is repeated until convergence. 
\end{algo}

\bigskip
When the original posteriors of $\balp$ and $\bbe$ do not have an enough mass around the region where the constraint holds, the importance weight could be degenerated, which would be a main drawback of Algorithm~1. 
To solve this issue, we also propose a sequential Monte Carlo algorithm to generate random samples from the tilted posterior with a sequence of parameters, $\{\lambda_{(0)}, \lambda_{(1)}, \ldots, \lambda_{(T)}\}$ with $\lambda_{(0)}=0$.
Note that the tilted posterior with $\lambda=\lambda_{(0)}$ reduces to the original posterior of $(\balp, \bbe)$, from which we can generate random samples. 
The detailed sampling steps are described as follows:

\begin{algo}[Tuning parameter selection via sequential Monte Carlo] \label{algo:SMC}
We first generate $S$ samples $(\balp_0^{(s)}, \bbe_0^{(s)}) \ (s=1,\ldots,S)$ from the original posterior, $p_n(\balp\mid D)p_n(\bbe\mid D)$ and set the uniform weight $w_0^{(s)}=1/S$.
Starting with the initial value $\lambda_{(0)}=0$ and $t=0$, repeat the following procedures for $t=1,\ldots,T$. 
\begin{enumerate}
\item 
(Updating weight) \ Given the particles $(\balp_{t-1}^{(s)}, \bbe_{t-1}^{(s)})$, update the weight as 
$$
w_t^{(s)}=\frac{\exp\{(\lambda_{(t)}-\lambda_{(t-1)})B_n(\balp_{t-1}^{(s)}, \bbe_{t-1}^{(s)})\}}
{\sum_{s'=1}^S\exp\{(\lambda_{(t)}-\lambda_{(t-1)})B_n(\balp_{t-1}^{(s')}, \bbe_{t-1}^{(s')})\}}
$$

\item 
(Resampling) \ Generate $(\balp_{Re}^{(s)}, \bbe_{Re}^{(s)})$ from the multinomial distribution on $(\balp_{t-1}^{(s)}, \bbe_{t-1}^{(s)}) \ (s=1,\ldots,S)$ according to the updated weight $w_t^{(s)}$.

\item 
(Smoothing) \ A new particle $(\balp_{t}^{(s)}, \bbe_{t}^{(s)})$ is defined as 
$$
(\balp_{t}^{(s)}, \bbe_{t}^{(s)}) = a  (\balp_{Re}^{(s)}, \bbe_{Re}^{(s)}) 
+ (1-a)(\bar{\balp}_{t-1}, \bar{\bbe}_{t-1}) + \ep^{(s)}, \ \ \ \ \ep^{(s)}\sim N(0, (1-a^2)\Sigma_{t-1}),
$$
where $a$ is a smoothing coefficient (e.g. $a=0.99$), $(\bar{\balp}_{t-1}, \bar{\bbe}_{t-1})$ is a mean vector at $t-1$, namely, $\bar{\balp}_{t-1}=S^{-1}\sum_{s=1}^S\balp_{t-1}^{(s)}$ and $\bar{\bbe}_{t-1}=S^{-1}\sum_{s=1}^S\bbe_{t-1}^{(s)}$, and $\Sigma_{t-1}$ is the variance-covariance matrix of $(\balp_{t-1}^{(s)}, \bbe_{t-1}^{(s)}) \ (s=1,\ldots,S)$.
We also set the weight as $w_t^{(s)}=1/S$. 

\item
(Evaluation of constraint) \ Compute $\bar{B}_{(t)}\equiv S^{-1}\sum_{s=1}^S B_n(\balp_t^{(s)}, \bbe_t^{(s)})$ and exit the loop if $|\bar{B}_{(t)}|$ is smaller than a tolerance value.  
\end{enumerate}
\end{algo}

The above method is based on the kernel smoothing updating of particles \citep{liu2001combined}.
A notable feature of the above algorithm requires generating random samples from the original posterior as initial particles. 
Also, it does not require evaluation of the value of original posterior, but it only needs the evaluation of the DR constraint $B_n(\balp, \bbe)$. 
When Algorithm~2 is terminated at $t=t_0$, $\lambda_{(t_0)}$ will be a desirable tilting parameter and $(\balp_{t}^{(s)}, \bbe_{t}^{(s)})$ with equal weights are samples from the tilted posterior. 

To specify a sequence of tilting parameters, $\{\lambda_{(0)}, \lambda_{(1)}, \ldots, \lambda_{(T)}\}$, we first compute the posterior mean of the constraint, $S^{-1}\sum_{s=1}^S B_n(\balp^{(s)}, \bbe^{(s)})$ under $\lambda=0$.
When the value is negative, the optimal $\lambda$ would be positive, so that we can set $\lambda_{(t)}=t\bar{\lambda}/T$ for some large $\bar{\lambda}>0$.
On the other hand, when the value is positive, we can set $\lambda_{(t)}=-t\bar{\lambda}/T$.

\section{Theoretical Properties} \label{sec:theorems}

\subsection{Posterior consistency}\label{sec4.1}
We here discuss the theoretical properties of the tilted posterior and the double robustness of the posterior mean (\ref{eq:ATE-PM}) under the tilted posterior.   
First, we show the behavior of the tilted posterior when the outcome model is correctly specified, as given in the following lemma. 
\begin{lem}\label{lem:dr1} Assuming regularity conditions 
in Appendix A. When the outcome model (\ref{gl1}) is correctly specified, it holds the following property under $n\to\infty$: 
$$
\iint\left|\pi_{n,\lambda}(\balp,\bbe)- p_{n}(\balp\mid D)p_{n}(\bbe\mid D)\right| d\balp d\bbe\stackrel{}{\to}0.
$$
\end{lem}

\medskip\noindent
The proof of Lemma~1 is given in Appendix B. 
Lemma~1 indicates that the tilting term in (\ref{eq:tilted-posterior}) automatically disappears when the (parametric) outcome model is correctly specified.
Hence, the posterior inference on the ATE (\ref{eq:ATE}) is not affected by the propensity score model, leading to consistency of the posterior mean (\ref{eq:ATE-PM}).

Furthermore, the posterior mean (\ref{eq:ATE-PM}) has a double robustness property as shown in the following theorem:

\begin{thm}\label{thm:dr2} Assuming regularity conditions 
in Appendix A. When either the outcome model (\ref{gl1}) or the propensity score model (\ref{gl2}) is correctly specified, then it holds under $n\to\infty$ that 
\begin{align}\label{thm1}
{\rm E}_{\pi_{n,\lambda}(\bbe)}\left[\frac{1}{n}\sum_{i=1}^{n}\left\{m_{1}(\bX_{i};\bbe)-m_{0}(\bX_{i};\bbe)\right\}\right]\stackrel{P}{\to}\tau.
\end{align}
\end{thm}

\medskip\noindent
The proof of Theorem~1 is provided in Appendix B. 
A key property for Theorem~1 is that the posterior mean (\ref{eq:ATE-PM}) can successfully utilize the information of the propensity model owing to entropic tilting, when the propensity model is correctly specified. 
Thus, inference based on $\pi_{n,\lambda}(\bbe)$ is valid in terms of posterior means when either the outcome model or the propensity score model is correctly specified.

\begin{rem}
For our proposed methodology, model correctness is related to the assumed (pseudo-) likelihood. For instance, when we assume the full likelihood for the outcome model, a fully correct specification of the likelihood function is necessary. Whereas, when we assume only the mean construction for the outcome model based on general Bayes, correctness of the outcome mean construction is sufficient. This point is different from traditional double robustness \citep{tsiatis2006semiparametric} and previous Bayesian DR methods \citep{saarela2016bayesian,luo2023semiparametric}.
\end{rem}
\begin{rem}
When both models are correctly specified, the posterior distribution after posterior coupling is the same as the posterior before coupling (from Lemma \ref{lem:dr1}). Therefore, the efficiency is derived from the outcome model, making it more efficient than traditional DR estimators. This point is a benefit of our proposed methodology and can be confirmed in the next simulation experiments.
\end{rem}

\subsection{Benefit of posterior coupling under correct propensity score model}\label{sec4.2}
Using the ET condition (\ref{eq:constraint}), our proposed method achieves double robustness.
Additionally, it may have the potential to improve the convergence rate of the posterior for the outcome regression model.
Intuitively, by applying the ET condition (\ref{eq:constraint}), the initial variability arising from the posterior distribution of the outcome model can be suppressed (becomes $0$ exactly), which in turn may improve the convergence rate of (\ref{eq:ATE-PM}).
This property is quite appealing when we adopt nonparametric or semiparametric Bayesian models for outcomes, under which the posterior convergence rate is typically slower than parametric models \citep[e.g.][]{ghosal2000convergence} and may produce inefficient posterior inference.  
This point is discussed in more detail in Appendix C and is demonstrated through simulation experiments.


\subsection{Comparison with existing methods}\label{sec4.3}
In our proposed method, the concept is similar to \cite{saarela2016bayesian} and \cite{luo2023semiparametric}. In both methods, moment conditions such as \eqref{eq:balancing} play an important role in constructing a DR estimator. Their approaches are primarily based on non-Bayesian perspectives, with efforts to harmonize them with Bayesian contexts.

In \cite{saarela2016bayesian}, the authors proposed a Bayesian bootstrap-based estimator.
Let $\xi = (\xi_1, \dots, \xi_n)$ denote the weights for the Bayesian bootstrap, with $\Xi \mid D \sim \mathrm{Dir}(1, \dots, 1)$.
Using the sampled weights $\xi$, a DR estimator weighted by $\xi$ is constructed:
$$
\sum_{i=1}^{n}\xi_{i}\left\{m_{1}(\bX_{i};\bbe(\xi))-m_{0}(\bX_{i};\bbe(\xi))+\frac{A_{i}-e(\bX_{i};\balp(\xi))}{e(\bX_{i};\balp(\xi))(1-e(\bX_{i};\balp(\xi)))}\left(Y_{i}-m_{A_{i}}(\bX_{i};\bbe(\xi))\right)\right\},
$$
where $\balp(\xi)$ and $\bbe(\xi)$ denote estimators obtained using the Bayesian bootstrap with $\xi$.
The target estimand is defined as
$$
{\rm E}_{\Xi\mid D}\left[{\rm E}_{\mathcal{E}}\left[Y\mid A=1,D,\xi\right]-{\rm E}_{\mathcal{E}}\left[Y\mid A=0,D,\xi\right]\right],
$$
and they show that the proposed estimator is consistent with this estimand in expectation when either the propensity score model or the outcome model is correctly specified.
Here, $\mathcal{E}$ represents the target population defined by the estimated propensity score. Roughly speaking, this population can be regarded as a ``random allocation" population. In general, this estimand is not consistent with the ATE $\tau$; the former can be considered a mixed ATE, whereas the latter corresponds to the (super)population ATE \citep{li2023bayesian}.

In \cite{yiu2020inference} and \cite{luo2023semiparametric}, a novel Bayesian method was proposed by tailoring the empirical likelihood framework \citep{owen2001empirical}, where the ATE $\tau$ is the target estimand.
Here, to clarify the discussion, we consider the true outcome model as
$$
Y_{i}=\tau A_{i}+h_{0}(\bX_i;\bbe^{0})+\varepsilon_{i},
$$
where $\varepsilon_{i}$ is an {\it i.i.d.}~error. To estimate the parameters, we consider the working model
$$
{\rm E}\left[Y_{i}\mid A_{i},\bX_{i}\right]=\tilde{\tau} A_{i}+h_{1}(\bX_i;\bbe)+\phi e_{i}(\hat{\balp}).
$$
The estimated propensity score is plugged into the model; however, simultaneous estimation can also be considered \citep{yiu2020inference}. To estimate the parameters, we consider the following estimating equation:
\begin{align}
\label{eeq_YL}
\sum_{i=1}^{n}U_{i}(\bbe, \phi, \tilde{\tau})=\sum_{i=1}^{n}\left(
\begingroup
\renewcommand{\arraystretch}{0.75}
\begin{array}{c}
A_{i}\\
\partial h_{1}(\bX_{i};\bbe)/\partial \bbe\\
e_{i}(\hat{\balp})
\end{array}
\endgroup
\right)\left(Y_{i}-\tilde{\tau} A_{i}-h_{1}(\bX_i;\bbe)-\phi e_{i}(\hat{\balp})\right).
\end{align}
By subtracting the 3rd row of (\ref{eeq_YL}) from the 1st row, we obtain
$$
\widehat{\tilde{\tau}}=\frac{\sum_{i=1}^{n}(A_{i}-e_{i}(\hat{\balp}))(Y_{i}-h_{1}(\bX_i;\hat{\bbe})-\hat{\phi} e_{i}(\hat{\balp}))}{\sum_{i=1}^{n}A_{i}(A_{i}-e_{i}(\hat{\balp}))},
$$
and this estimator is related to the DR estimator (\ref{dr1}). This is because, from the true outcome model,
$$
\widehat{\tilde{\tau}}\approx\tau+\frac{\sum_{i=1}^{n}(A_{i}-e_{i}(\hat{\balp}))(h_{0}(\bX_i;\bbe^{0})-h_{1}(\bX_i;\hat{\bbe})-\hat{\phi} e_{i}(\hat{\balp}))}{\sum_{i=1}^{n}A_{i}(A_{i}-e_{i}(\hat{\balp}))}.
$$
The estimator has a form similar to (\ref{dr1}), and the second term clearly converges to zero asymptotically when either the propensity score or a part of the outcome model (i.e., $h_{1}(\bX_i;\bbe)$) is correctly specified.
Using (\ref{eeq_YL}), we can consider the empirical likelihood estimator. Specifically, the estimated empirical likelihood function can be expressed as
$$
p_{i}(\bbe,\phi,\tilde{\tau})=\frac{\exp\left\{\hat{\lambda}^{\top}U_{i}(\bbe,\phi,\tilde{\tau})\right\}}{\sum_{j=1}^{n}\exp\left\{\hat{\lambda}^{\top}U_{j}(\bbe,\phi,\tilde{\tau})\right\}},
$$
where $\lambda$ is a Lagrange multiplier. Using $p_{i}$, we can consider the posterior distribution
$$
\pi(\bbe,\phi,\tilde{\tau}\mid D)\propto\prod_{i=1}^{n}p_{i}(\bbe,\phi,\tilde{\tau})\pi(\bbe,\phi,\tilde{\tau}).
$$
Based on the posterior distribution, one can obtain samples that possess the double robustness property. The origin of $\lambda$ differs from that in our proposed method; however, in both their methods and our proposed approach, the estimation (or adjustment) of $\lambda$ plays an essential role in achieving double robustness. Compared with this procedure, we consider our proposed methodology more straightforward, as it simply combines the posterior distributions for the propensity score and outcome models.



\section{Simulation experiments} \label{sec:sim}
To confirm performance of our proposed procedure, we conducted simulation experiments. The iteration time of all simulations was 2000. Some additional simulation experiments are presented in Appendix D.

\subsection{Data-generating mechanism}\label{sec5.1}
The data-generating mechanism was based on the setting of \cite{Ka2007}. 
We describe the data-generating mechanism used in the simulations. Assume that there were four covariates, denoted as $\bX_{i}=(X_{1i},X_{2i},X_{3i},X_{4i})$. Each $X_{ji}$ was independently generated from the standard normal distribution. Next, we introduce the assignment mechanism for the treatment value $A_{i}$; specifically, the true propensity score was defined as
$$
e(\bX_{i})={\rm Pr}\left(A=1 \mid \bX_{i}\right)={\rm expit}\left\{X_{1i}-0.5 X_{2i}+0.25 X_{3i}+0.1 X_{4i}\right\}. 
$$
Finally, we introduce the model for the potential outcomes
\[
Y_{ai}=100+110a+13.7 \times (2 X_{1i}+X_{2i}+X_{3i}+X_{4i})+\varepsilon_{i},
\]
where $\varepsilon_{i}$ was generated from the standard normal distribution. Under these settings, the ATE was $\Delta_{0}={\rm E}[Y_{1}-Y_{0}]=110$.

\subsection{Estimating methods and performance metrics}\label{sec5.2}
We compared four methods:\ one non-Bayesian DR estimator proposed by \cite{Bang2005doubly}, one Bayesian G-formula based method \citep{daniels2023bayesian}, two Bayesian DR estimator proposed by \cite{saarela2016bayesian} and \cite{luo2023semiparametric}, and the proposed Bayesian DR estimator using posterior coupling.

To evaluate the four methods, we consider three situations:\ 1) both the propensity score and outcome model is correctly specified, 2) only the propensity score model is correctly specified, and 3) only the outcome model is correctly specified. For misspecified model,
only covariate $X_1$ is used for each model.

We evaluated the various methods based on mean, empirical standard error (ESE), root mean squared error (RMSE), coverage probability (CP), average length of confidence / credible intervals, and boxplot of estimated ATE from $J=2000$ replications. The RMSE were calculated as ${\rm RMSE}=\sqrt{J^{-1}\sum_{j=1}^J(\hat{\Delta}_{j}-\Delta_{0})^2}$, where $\hat{\Delta}_{j}$ is the estimate of each estimator at $j$th replication, and $\Delta_{0}$ ($=110$) is the true value of the ATE. The CP refers to the proportion of cases where the 95\% confidence / credible interval includes $\Delta_{0}$.

\subsection{Simulation results}\label{sec5.3}
\begin{figure}[htbp]
\begin{center}
\begin{tabular}{c}
\includegraphics[width=16.5cm]{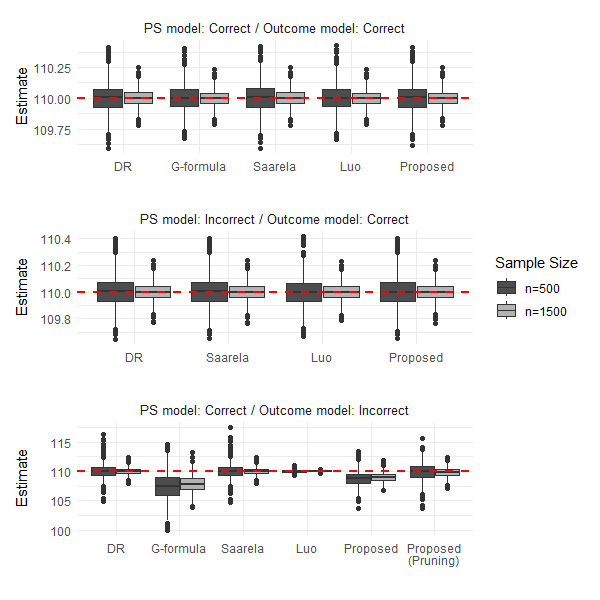}
\end{tabular}\caption{Boxplot of causal effect estimates:\ The number of iteration is $2000$ and the true value is 110 (red dashed line).}
\label{fig1}
\end{center}
{\footnotesize{Correct:\ propensity score / outcome model is correctly specified; Incorrect:\ propensity score / outcome model is misspecified.\\
DR:\ Ordinaly non-Bayesian DR estimator that is asymptotically equivalent to \cite{Bang2005doubly}; 
G-formula:\ Bayesian G-formula based method discussed in \cite{daniels2023bayesian}; 
Saarela:\ Bayesian DR estimator using Bayesian Bootstrap method proposed by \cite{saarela2016bayesian}; 
Luo:\ Bayesian DR estimator using empirical likelihood-based method proposed by \cite{luo2023semiparametric}.\\
For our proposed method, sequential Monte Carlo with replacement is conducted. For all Bayesian methods, 20,000 posterior draws are used.\\
Note:\ The range of the y-axis differs for each figure.
}}
\end{figure}

The results are summarized in Table \ref{tab1} and Figure \ref{fig1}. When both the propensity score and outcome models are correctly specified, the DR and Saarela's methods exhibit nearly identical performance. This result implicitly shows that the DR and Saarela's methods achieve the semiparametric efficiency bound \citep{tsiatis2006semiparametric}. Meanwhile, the G-formula, Luo's method, and our proposed method achieve better ESE compared to the DR methods. This is an attractive point, as our proposed DR method is potentially more efficient than ordinary DR methods.

When only the outcome model is correctly specified, the proposed method shows performance nearly comparable to that of the G-formula. This result is consistent with Lemma 1. The DR and Saarela’s methods again show similar performance. Even in this situation, the G-formula, Luo's method, and our proposed method achieve better ESE compared to the DR methods.
On the other hand, when only the propensity score model is correctly specified, the bias of the proposed method is improved compared to that of the G-formula under both small and large sample situations. This result is consistent with Theorem 1. Additionally, the ESE, RMSE, and CP are improved. The DR and Saarela’s methods exhibit almost similar performance, but the CP shows different results. Luo's method also has good performance; however, the CP is too conservative. This tendency is similar to that observed for the DR.

From these results, our proposed method demonstrates the DR property while achieving fully Bayesian inference. When only the propensity score model is correctly specified, the bias is improved; however, some residual bias remains. Therefore, the specification of the outcome model is more important compared to that of the propensity score model.

\subsubsection{Remaining bias modification}
As mentioned in the previous section, when only the propensity score model is correctly specified, the proposed method exhibits smaller bias compared to the G-formula. However, some bias still remains. This issue is related to the violation of condition (C.3) in Appendix A. In large sample settings, since $B_{n}\stackrel{P}{\not\to}0$, some samples must carry large sampling weights. To accommodate this, the parameter $\lambda$ becomes large, which leads to a violation of (C.3).

To address this problem, we propose the ``sample pruning" algorithm. When updating $\lambda$ in the SMC algorithm, we discard samples with small sampling weights (i.e., $\exp\left\{\lambda B_{n}\right\}$). As a result, the remaining samples are more concentrated around $B_{n}\approx0$ without extreme sampling weights. The results of the sample pruning algorithm are presented in the last row of Table \ref{tab1}. The remaining bias is clearly diminished.

[Table \ref{tab1} and Figure \ref{fig1} are here.]

\subsection{Simulation results using BART}\label{sec5.4}
As mentioned in Section \ref{sec4.2}, the proposed method can improve the convergence rate of nonparametric methods. The results are summarized in Table \ref{tab2}. As expected, the proposed method improves upon the bias and variance of the results of the G-formula, especially in small-sample situations. This is because our proposed method balances the moment condition (\ref{eq:balancing}) using ET, and this term diminishes (goes to $0$) asymptotically. Additionally, the simulation results suggest that using ET improves efficiency even when the propensity score model is misspecified.

These results suggest that detecting a valid outcome model is an initially important task, and incorporating a propensity score model (information) may help improve the efficiency of the ATE estimation.

[Table \ref{tab2} is here.]

\section{Application 1: Effects of Antihypertensive Treatment on Dementia and  Sensitivity Analysis} \label{sec:sensitivity}

\subsection{Background}\label{sec6.1}
Antihypertensive treatments have been shown to reduce the risk of dementia in previous studies \citep{marpillat2013antihypertensive,ding2020antihypertensive,bennett2025target}.
Among these, we highlight \cite{bennett2025target} as a motivating example.
As emphasized in their analysis, although adjustments were made for observed confounders, the possibility of important unmeasured confounders remains.
Here, our aim is to assess the potential impact of such unmeasured confounders through sensitivity analysis, which can be readily incorporated into the posterior coupling framework.

We use the HRS data available from the Health and Retirement Study website (\url{https://hrsdata.isr.umich.edu/data-products/}). For our analysis, data from 2010 to 2016 were used at two-year intervals. Specifically, we defined exposure and baseline characteristics using the 2010 and 2012 surveys, and assessed outcome variables using the data collected in the survey conducted four years later. Subjects who had dementia, Alzheimer's disease, memory problems, or who received any antihypertensive treatment at baseline were excluded from the analysis. Additionally, subjects aged 65 years or older with systolic blood pressure greater than 130 at baseline were included in the analysis. The exposure variable was defined as ``using any antihypertensive treatment and having hypertension" or not. The outcome variable was defined as ``dementia, Alzheimer's disease, or memory problems" or not. Confounders were selected based on \cite{bennett2025target}:\ baseline SBP, age, sex, education, race, body mass index, apolipoprotein E $\varepsilon$4 allele status, smoking, depression, coronary heart disease, stroke, and diabetes. Under these definitions, we can evaluate the effect of antihypertensive use on dementia incidence.

\subsection{Sensitivity analysis of unmeasured confounding}\label{sec6.2}
We consider Bayesian sensitivity analysis methods \citep{mccandless2007bayesian,daniels2023bayesian,zou2025bayesian}; specifically focusing on the approach discussed in \cite{daniels2023bayesian}. In their framework, the contrast in expectations without assuming strongly ignorable treatment assignment (i.e., in the presence of unmeasured confounding) is expressed as
$$
{\rm E}\left[{\rm E}\left[Y\mid A=1,\bX\right]-{\rm E}\left[Y\mid A=0,\bX\right]\right]=\tau+\xi,
$$
where
$$
\xi={\rm E}\left[\Delta_{0}(\bX){\rm Pr}(A=1\mid \bX)+\Delta_{1}(\bX){\rm Pr}(A=0\mid \bX)\right],
$$
and $\Delta_{a}(\bX)={\rm E}\left[Y_{a}\mid A=1,\bX\right]-{\rm E}\left[Y_{a}\mid A=0,\bX\right]$ (more details, \citealt{hu2022flexible}). Note that the assumption of strongly ignorable treatment assignment corresponds to $\xi = 0$. A simple but important case is when $\Delta_a(\bX) \equiv \Delta$, which corresponds to the situation where $\bX$ provides no information about unmeasured confounders \citep{daniels2023bayesian}. In this case, $\xi = \Delta$, and $\xi$ serves as the sole sensitivity parameter.

To consider the sensitivity parameter, we adopt an outcome model, $m_{A}(\bX;\bbe,\xi)=m_{A}(\bX;\bbe)+A\xi$. Let $f(Y_i\mid A_i,\bX_i;\bbe,\xi_{i})$ be the corresponding negative log-likelihood function of the outcome model including $\xi_{i}$. 
Then, the joint posterior of $(\balp, \bbe)$ is proportional to 
\begin{equation}\label{eq:pos-sensitivity}
\pi(\balp)\pi(\bbe)\prod_{i=1}^n \exp\left\{- f(A_i\mid \bX_i;\balp) \right\} \int \exp\left\{-f(Y_i\mid A_i,\bX_i;\bbe,\xi_{i}) \right\}g(\xi_{i})d\xi_{i},
\end{equation}
where $g(\xi_{i})$ is a density function of $\xi_{i}$.
Then, by using the posterior samples from the original posterior $p_n(\balp,\bbe\mid D)$ with the base model $m_{A}(\bX;\bbe)$, the posterior samples from the above posterior can be generated by an importance sampling with important weight proportional to 
$$
\prod_{i=1}^n \frac{\int \exp\left\{-f(Y_i\mid A_i,\bX_i;\bbe,\xi_{i})\right\}g(\xi_{i})d\xi_{i}}{\exp\left\{-f(Y_i\mid A_i,\bX_i;\bbe)\right\}}.
$$
Assuming that random sample generation from $g(\cdot)$ is tractable, we propose the following procedure to obtain the posterior sample from (\ref{eq:pos-sensitivity}).

\begin{algo} (Importance sampling approach for sensitivity analysis)
For $s=1,\ldots,S$, generate $M$ samples from $g(\cdot)$, denoted by $\xi^{(m)}_{i}$ for $m=1,\ldots,M$ and compute the weight as 
$$
w^{(s)}\equiv \prod_{i=1}^n \frac{M^{-1}\sum_{m=1}^M\exp\left\{-f(Y_i\mid A_i,\bX_i;\bbe^{(s)},\xi^{(m)}_{i})\right\}}{\exp\left\{-f(Y_i\mid A_i,\bX_i;\bbe^{(s)})\right\}}.
$$
Then, the posterior samples can be approximated by the particles $(\balp^{(s)},\bbe^{(s)})$ and its weight $w^{(s)}$ for $s=1,\ldots,S$.
\end{algo}
\noindent
Note that in the above sensitivity analysis case ($\xi = \Delta$), $\xi_{i}$ can be considered as $\xi_{i}\equiv\xi$. Another importance weighting derivation strategy is discussed in Appendix \ref{app:impwgt_kuan}.

Given the posterior samples of $(\balp, \bbe)$ under unmeasured confounders, we can also obtain the DR posterior samples of the ATE through posterior coupling, described in Algorithms~\ref{algo:IS} or \ref{algo:SMC}.

\subsection{Results}\label{sec6.3}
Without adjusting for these observed confounders (i.e., crude analysis), the point estimate of the risk difference and its 95\% CI is $-0.011$ $(-0.039, 0.018)$. The posterior mean of the ATE and its 95\% CI using the G-formula is $0.007$ $(-0.114, 0.127)$. For our proposed method, the posterior mean of the ATE and its 95\% CI is $0.010$ $(-0.111, 0.130)$.

In line with \cite{bennett2025target}, we examine the impact of unmeasured confounders on the analysis results. Specifically, we conduct the sensitivity analysis described in Section \ref{sec6.2}. To do so, the sensitivity parameter must be set carefully. As an example, we specify two distributions: one is a triangular distribution on $(0,0.5)$ for $\xi$ with a peak at $0.5$, and the other is a triangular distribution on $(-0.5,0)$ with a peak at $-0.5$. These distributions represent scenarios where unmeasured confounders have positive/negative impacts on the causal effects. For the former case, the posterior mean of the ATE and its 95\% CI is $0.002$ $(-0.097, 0.143)$, whereas for the latter case, it is $-0.004$ $(-0.122, 0.109)$. These results raise some concerns regarding unmeasured confounders; however, their impact is unlikely to alter the interpretation of the main analysis results.

\section{Application 2: Impact of Right Heart Catheterization on Mortality with Confounder Selection} \label{sec:selection}

\subsection{Background}\label{sec7.1}
It is recognized that Right Heart Catheterization (RHC) has a fatal impact even after adjusting for sufficient confounders \citep{connors1996effectiveness, chen2020right}.
Here, we use the dataset available from the website:\ \url{https://hbiostat.org/data/}. The exposure variable is defined as RHC or not, and the outcome variable is death within 30 days after admission. In line with \cite{harada2025false}, we select confounders from 48 potential candidates.
A serious practical issue is that the inclusion or exclusion of confounders can substantially affect causal conclusions: incorporating irrelevant variables may inflate variance, while omitting important ones can lead to bias.
To mitigate this issue, we propose a confounder selection approach similar to the method by \cite{ning2020robust}, which is incorporated into the posterior coupling framework.

\subsection{Confounder selection}\label{sec7.2}
We consider applying shrinkage techniques to both the propensity score and outcome models. A straightforward implementation is to apply shrinkage methods separately to each model. However, such a ``simple procedure" may overlook weak confounders, potentially leading to regularization-induced confounding \citep{hahn2018regularization}. To address this issue, we introduce a novel Bayesian method, which can be seen as a modification of the sequential confounder selection approach proposed by \cite{ning2020robust}.

\begin{algo}[Confounder selection with posterior coupling]\phantom{a}
\label{algo:selection}
\begin{itemize}
\item
(Step 1) Generate posterior samples for the propensity score and outcome models using shrinkage priors, such as the horseshoe prior \citep{Ca2010}. Additionally, define the subset of selected covariates as $S=\{j:\balp_{j}\neq0\}$, where $\balp_{S}$ denotes the coefficients associated with the selected predictors for the propensity score model.
\item
(Step 2) Modify the moment condition in (\ref{eq:balancing}) as
$$
B^{S}_{n}(\balp_{S},\bbe_{S})\equiv \frac1n\sum_{i=1}^n \frac{A_{i}-e(\bX_{Si};\balp_{S})}{e(\bX_{Si};\balp_{S})(1-e(\bX_{Si};\balp_{S}))}\left(Y_{i}-m_{A_{i}}(\bX_{i};\bbe_{S},\bbe_{S^{c}})\right),
$$
and obtain posterior samples from the tilted posterior using Algorithms~\ref{algo:IS} or \ref{algo:SMC}, where only $(\balp_{S}, \bbe_{S})$ are updated with $\bbe_{S^{c}}$ being unchanged.
\end{itemize}
\end{algo}

The above selection procedure has several advantages.
First, previously overlooked confounders may have a chance to be revived through the use of posterior coupling. 
Second, by focusing the update on important covariates (i.e., confounders), the proposed posterior coupling algorithm such as SMC (Algorithm~\ref{algo:SMC}) can work more effectively. 
Finally, when the outcome model is correctly specified, the posterior coupling does not interfere with the double robustness property. Therefore, when there are a large number of confounders, Algorithm~\ref{algo:selection} is expected to improve the accuracy of causal effect estimation.


\subsection{Results}\label{sec7.3}
To estimate the ATE (i.e., the risk difference), we applied Algorithm 4 described in Section \ref{sec7.2}. In this analysis, we define the subset of selected covariates as $S=\{j:|\bar{\balp}_{j}|\geq 0.01\}$, where $\bar{\balp}_{j}$ denotes the posterior mean in Step 1 of Algorithm 4 for the $j$-th covariate. From Step 1, 44 of the 58 covariates are selected. Note that there are 58 candidates because dummy variables are included. The posterior mean of the ATE and its 95\% CI for the G-formula is $0.229$ $(0.032, 0.395)$, with a CI length of $0.363$. Without applying the algorithm (i.e., without covariate selection), the posterior mean of the ATE and its 95\% CI is $0.043$ $(-0.075, 0.205)$, with a CI length of $0.280$. In contrast, with the algorithm, the estimates are $0.188$ $(0.122, 0.312)$, with a CI length of $0.190$. These results suggest that applying the algorithm allows more accurate estimation of causal effects, which is consistent with the discussion in Appendix \ref{add_sim_HD}. Based on these findings, we recommend that updating the selected posterior with ET is necessary to accurately estimate causal effects when there are many potential confounders.



\section{Discussion} \label{sec:discussion}
In this manuscript, we propose a novel Bayesian DR estimator whose posterior distribution can be described explicitly. Our proposed method achieves this by using an entropic tilting condition, which is related to the DR estimator proposed by \cite{Bang2005doubly}. This condition plays a role in modifying the posterior distribution for the outcome model by incorporating information from the propensity score model. As shown in both the mathematical discussions and the simulation results, our proposed method exhibits double robustness. The explicit posterior formulation also enables important extensions such as sensitivity analysis for unmeasured confounding and confounding selection in high-dimensional settings.

As mentioned in the Introduction, many Bayesian DR estimators have been proposed \citep{saarela2016bayesian, yiu2020inference, antonelli2022causal, breunig2025double}. However, an explicit description of the posterior distribution (i.e., prior distributions and likelihoods) is particularly attractive. 
The explicit posterior formulation enables various flexible estimation strategies, including hierarchical Bayesian modeling for clustered data---a structure appealing in several fields, including biostatistics and medicine \citep{yang2018propensity, chang2022propensity}. Additionally, combining dynamic treatment regime settings with Bayesian DR inference is attractive \citep{dong2025estimating}.
Furthermore, as mentioned in Appendix \ref{app:subclass}, the posterior description enables the construction of an algorithm for estimating the optimal number of strata for propensity score subclassification \citep{orihara2024bayesian}.

\vspace{0.5cm}

\noindent
{\bf Acknowledgement:} We have prepared an R package for our proposed methodology. For more details, see the GitHub page of Dr.\ Momozaki:\ \url{https://github.com/t-momozaki/DRBayes}. We would like to express our gratitude to Dr.\ Kosuke Inoue, Dr.\ Kuan Liu, and Dr.\ Yu Luo for their comments and support. Dr.\ Liu also provided her idea for the discussion in Appendix \ref{app:impwgt_kuan}.

\vspace{0.2cm}
\noindent
{\bf Funding:} This work was supported by JSPS KAKENHI Grant Numbers 24K21420, 25K21166, and 25H00546.

\vspace{0.2cm}
\noindent
{\bf Conflict of interest:} The authors declare no conflicts of interest.

\vspace{0.2cm}
\bibliographystyle{chicago}
\bibliography{ref}

\newpage

\begin{landscape}
\vspace{-3cm}
\begin{table}[h]
\begin{center}
\caption{Summary of causal effect estimates:\ The number of iteration is $2000$ and the true ATE is 110. The absolute bias (ABias), empirical standard error (ESE), root mean squared error (RMSE), coverage probability (CP), and average length of credible interval (AvL) of the estimated causal effects across 2000 iterations are summarized by propensity score model specification (``PS model" column), outcome model specification (``Outcome model" column), and estimation method (``Method" column).}
\label{tab1}
\begin{tabular}{ccc|ccccc|ccccc}\hline
{\bf PS}&{\bf Outcome}&{\bf Method}&\multicolumn{5}{|c}{\bf Sample size:\ $n=500$}&\multicolumn{5}{|c}{\bf Sample size:\ $n=1500$}\\\cline{4-13}
{\bf model}&{\bf model}& & ABias & ESE & RMSE & CP & AvL & ABias & ESE & RMSE & CP & AvL \\ \hline
Correct & Correct & DR        & 0.002 & 0.113 & 0.113 & 93.0 & 0.420 & 0.000 & 0.064 & 0.064 & 95.0 & 0.244 \\ \cline{3-13}
& & G-formula & 0.001 & 0.105 & 0.105 & 93.6 & 0.397 & 0.001 & 0.060 & 0.060 & 94.7 & 0.228 \\ \cline{3-13}
& & Saarela   & 0.002 & 0.113 & 0.113 & 92.5 & 0.413 & 0.000 & 0.064 & 0.064 & 94.8 & 0.243 \\ \cline{3-13}
& & Luo       & 0.001 & 0.106 & 0.106 & 93.1 & 0.390 & 0.001 & 0.060 & 0.060 & 94.6 & 0.226 \\ \cline{3-13}
& & Proposed  & 0.001 & 0.110 & 0.110 & 92.2 & 0.396 & 0.001 & 0.061 & 0.061 & 94.2 & 0.227 \\ \hline
Incorrect & Correct & DR      & 0.001 & 0.108 & 0.108 & 92.7 & 0.392 & 0.001 & 0.062 & 0.062 & 94.2 & 0.228 \\ \cline{3-13}
& & Saarela   & 0.001 & 0.108 & 0.108 & 93.3 & 0.401 & 0.001 & 0.062 & 0.062 & 94.8 & 0.234 \\ \cline{3-13}
& & Luo       & 0.001 & 0.106 & 0.106 & 93.0 & 0.390 & 0.001 & 0.060 & 0.060 & 94.6 & 0.226 \\ \cline{3-13}
& & Proposed  & 0.000 & 0.108 & 0.108 & 92.8 & 0.396 & 0.001 & 0.061 & 0.061 & 94.2 & 0.227 \\ \hline
Correct & Incorrect & DR      & 0.016 & 1.025 & 1.025 & 100 & 10.070 & 0.013 & 0.568 & 0.568 & 100 & 5.826 \\ \cline{3-13}
& & G-formula & 2.535 & 2.286 & 3.414 & 80.8 & 8.969 & 2.108 & 1.315 & 2.485 & 63.4 & 5.208 \\ \cline{3-13}
& & Saarela   & 0.008 & 1.056 & 1.056 & 92.2 & 3.807 & 0.012 & 0.572 & 0.572 & 92.4 & 2.135 \\ \cline{3-13}
& & Luo       & 0.004 & 0.205 & 0.205 & 100 & 9.094 & 0.002 & 0.104 & 0.104 & 100 & 5.308 \\ \cline{3-13}
& & Proposed  & 1.267 & 1.271 & 1.795 & 98.9 & 8.960 & 1.068 & 0.735 & 1.296 & 97.0 & 5.186 \\ \hline\hline

Correct & Incorrect & \begin{tabular}{c}{Proposed} \\ {(pruning)} \end{tabular} & 0.086 & 1.428 & 1.430 & 98.0 & 7.539 & 0.154 & 0.777 & 0.792 & 98.4 & 4.340 \\ \hline
\end{tabular}
\end{center}
{\footnotesize{Correct:\ propensity score / outcome model is correctly specified; Incorrect:\ propensity score / outcome model is misspecified.\\
DR:\ Ordinaly non-Bayesian DR estimator that is asymptotically equivalent to \cite{Bang2005doubly}; 
G-formula:\ Bayesian G-formula based method discussed in \cite{daniels2023bayesian}; 
Saarela:\ Bayesian DR estimator using Bayesian Bootstrap method proposed by \cite{saarela2016bayesian}; 
Luo:\ Bayesian DR estimator using empirical likelihood-based method proposed by \cite{luo2023semiparametric}; Pruning:\ Using sample pruning algorithm for our proposed method described in Section \ref{sec5.3}.1\\
For our proposed method, sequential Monte Carlo with replacement is conducted. For all Bayesian methods, 20,000 posterior draws are used.
}}
\end{table}

\end{landscape}

\newpage
\begin{table}[h]
\begin{center}
\caption{Summary of causal effect estimates under BART model:\ The number of iteration is $2000$ and the true ATE is 110. The absolute bias (ABias), empirical standard error (ESE), and root mean squared error (RMSE) of the estimated causal effects across 2000 iterations are summarized by estimation method (``Method" column).}
\label{tab2}
\scalebox{0.9}{
\begin{tabular}{c|ccc|ccc|ccc}\hline
{\bf Method}&\multicolumn{3}{|c}{\bf Sample size:\ $n=200$}&\multicolumn{3}{|c}{\bf Sample size:\ $n=500$}&\multicolumn{3}{|c}{\bf Sample size:\ $n=1500$}\\\cline{2-10}
& ABias & ESE & RMSE & ABias & ESE & RMSE & ABias & ESE & RMSE \\ \hline
G-formula & 0.428 & 0.949 & 1.042 & 0.189 & 0.365 & 0.410 & 0.076 & 0.139 & 0.159 \\ \hline
\begin{tabular}{c}{Proposed} \\ {(PS correct)} \end{tabular} & 0.424 & 0.921 & 1.014 & 0.183 & 0.355 & 0.400 & 0.067 & 0.138 & 0.153 \\ \hline
\begin{tabular}{c}{Proposed} \\ {(PS incorrect)} \end{tabular} & 0.422 & 0.921 & 1.013 & 0.181 & 0.355 & 0.398 & 0.066 & 0.138 & 0.153 \\ \hline
\end{tabular}
}
\end{center}
{\footnotesize{
G-formula:\ Bayesian G-formula based method discussed in \cite{daniels2023bayesian}.\\
PS correct:\ propensity score model is correctly specified; PS incorrect:\ propensity score model is misspecified.\\
For our proposed method, importance sampling without replacement is conducted. For all methods, 10,000 samples out of 20,000 posterior draws are used.
}}
\end{table}

\appendix

\newpage
\section{Regularity conditions}
\begin{description}
\item[C.1]
For arbitrary $\lambda$, it holds that 
$$
\sup_{\balp\in\Theta_{\balp},\ \bbe\in\left(A_{\varepsilon}^{\bbe}\right)^{c}}\left|1-\exp\left\{\lambda B_n(\balp,\bbe)\right|\right|<\infty.
$$

\item[C.2] 
For some convergence point $\balp^{*}\in\Theta_{\balp}$ and the true value $\bbe^{0}\in\Theta_{\bbe}$ (i.e., ${\rm E}\left[Y\mid A_{i},\bX_{i}\right]=m_{A_{i}}(\bX_{i};\bbe^{0})$), it holds that 
$$
\left|\left|\frac{\partial}{\partial\bbe}B_n(\balp^{*},\bbe^{0})\right|\right|=\frac1n \left|\left|\sum_{i=1}^n\frac{A_{i}-e(\bX_{i};\balp^{*})}{e(\bX_{i};\balp^{*})(1-e(\bX_{i};\balp^{*}))}\left(\frac{\partial}{\partial\bbe}m_{A_{i}}(\bX_{i};\bbe^{0})\right)\right|\right|<\infty.
$$
\item[C.3]
For $\balp\in \Theta_{\balp}$ and $\bbe\in \Theta_{\bbe}$, it holds that 
\begin{align*}
&\int\left|\left|\frac{1}{n}\sum_{i=1}^{n}\left|\left(\frac{A_{i}}{e(\bX_{i};\balp)^2}+\frac{1-A_{i}}{(1-e(\bX_{i};\balp))^2}\right)\left(Y_{i}-m_{A_{i}}(\bX_{i};\bbe)\right)\left(\frac{\partial}{\partial\balp^{\top}}e(\bX_{i};\balp)\right)\right|\right.\right.\\
&\hspace{0.5cm}
\vphantom{\left|\frac{A_{i}}{e(\bX_{i};\balp)}\right|^2}   
\times \left.\left.\frac{\exp\left\{\lambda B_n(\balp,\bbe)\right\}}{z_{n}^{\balp,\bbe}}\right|\right|^2p_{n}(d\balp|D)p_{n}(d\bbe|D)<\infty.
\end{align*}
\end{description}
From condition {\bf (C.1)}, it is expected that $\balp$, $\bbe$, and $\lambda$ need to have compact support. Condition {\bf (C.2)} is a mild condition compared to other regularity conditions. Condition {\bf (C.3)} is difficult to interpret and can be regarded as purely a technical regularity condition.

\section{Proofs}
To complete the proofs, we introduce the following lemma.
\begin{alem}
\begin{align*}
{\rm E}_{p_{n}(\balp\mid D)}\left[\balp\right]\to \balp^{0},\ {\rm E}_{p_{n}(\balp\mid D)}\left[||\balp-\balp^{0}||^2\right]\to 0,\ \ \ and\\
{\rm E}_{p_{n}(\bbe\mid D)}\left[\bbe\right]\to \bbe^{0},\ {\rm E}_{p_{n}(\bbe\mid D)}\left[||\bbe-\bbe^{0}||^2\right]\to 0.
\end{align*}
\end{alem}

\subsection{Proof of Lemma 1}
First, define $\widetilde{\pi}_{n,\lambda}(\balp,\bbe)\equiv \exp\left\{\lambda B_n(\balp,\bbe)\right\}p_{n}(\balp|D)p_{n}(\bbe|D)$ as the unnormalized tilted posterior. 
Then, we have 
$$
\left|\widetilde{\pi}_{n,\lambda}(\balp,\bbe)-p_{n}(\balp|D)p_{n}(\bbe|D)\right|=\left|1-\exp\left\{\lambda B_n(\balp,\bbe)\right\}\right|p_{n}(\balp|D)p_{n}(\bbe|D).
$$
We evaluate the above difference separately for the following four subsets:
$$
R_1=(A_{\varepsilon}^{\balp})^c\times (A_{\varepsilon}^{\bbe})^c, \ \ \ \ 
R_2=A_{\varepsilon}^{\balp}\times (A_{\varepsilon}^{\bbe})^c,\ \ \ \ 
R_3=A_{\varepsilon}^{\balp}\times A_{\varepsilon}^{\bbe}, \ \ \ \
R_4=(A_{\varepsilon}^{\balp})^c\times A_{\varepsilon}^{\bbe}, \ \ \ \
$$
First, for $R_1$, it follows that 
\begin{align}
&\int_{R_1}\left|\widetilde{\pi}_{n,\lambda}(\balp,\bbe)-p_{n}(\balp|D)p_{n}(\bbe|D)\right|d\balp d\bbe\nonumber\\
&=\int_{R_1}\left|1-\exp\left\{\lambda B_n(\balp,\bbe)\right\}\right|p_{n}(\balp|D)p_{n}(\bbe|D)d\balp d\bbe\nonumber\\
&=\int_{R_1}\left|1-\exp\left\{\lambda B_n(\balp,\bbe)\right\}\right|\exp\left\{-n(f_{n}(\balp)+f_{n}(\bbe))\right\}p(\balp)p(\bbe)d\balp d\bbe\label{eql1}\\
&\leq\left\{\sup_{\balp\in\left(A_{\varepsilon}^{\balp}\right)^{c},\ \bbe\in\left(A_{\varepsilon}^{\bbe}\right)^{c}}\left|1-\exp\left\{\lambda B_n(\balp,\bbe)\right\}\right|\right\} \\
& \quad 
\times \int_{\balp\in (A_{\varepsilon}^{\balp})^{c}} \exp\left\{-nf_{n}(\balp)\right\}p(\balp)d\balp
\int_{\bbe\in (A_{\varepsilon}^{\bbe})^{c}}\exp\left\{-nf_{n}(\bbe)\right\}p(\bbe) d\bbe\nonumber.
\end{align}
From the regularity condition of \cite{miller2021asymptotic}, the second and third term becomes $0$. Therefore, from {\bf (C.1)}, the integral discussed above becomes $0$. 
The same argument holds for $R_2$. 

For $R_3$, from (\ref{eql1}),
\begin{align}
&\int\left|1-\exp\left\{\lambda B_n(\balp,\bbe)\right\}\right|\exp\left\{-n(f_{n}(\balp)+f_{n}(\bbe))\right\}p(\balp)p(\bbe)d\balp d\bbe\nonumber\\
&<\left\{\sup_{\balp\in A_{\varepsilon}^{\balp},\ \bbe\in A_{\varepsilon}^{\bbe}}\left|1-\exp\left\{\lambda B_n(\balp,\bbe)\right\}\right|\right\}\times(1+\varepsilon)^2.\label{eql2}
\end{align}
Here, the last inequality becomes Proposition 1. From Taylor expansion for the first term of the above inequality around $(\balp^{*},\bbe^{0})$,
\begin{align*}
\exp\left\{\lambda B_n(\balp,\bbe)\right\}=1+\lambda B_n(\balp^{*},\bbe^{0})+\frac{\partial}{\partial\balp^{\top}}B_n(\balp^{*},\bbe^{0})(\balp-\balp^{*})+\frac{\partial}{\partial\bbe^{\top}}B_n(\balp^{*},\bbe^{0})(\bbe-\bbe^{0})
\end{align*}
under sufficient large $n$.
When the outcome model is correctly specified,
\begin{align*}
B_n(\balp^{*},\bbe^{0})&=\frac{1}{n}\sum_{i=1}^n \frac{A_{i}-e(\bX_{i};\balp^{*})}{e(\bX_{i};\balp^{*})(1-e(\bX_{i};\balp^{*}))}\left(Y_{i}-m_{A_{i}}(\bX_{i};\bbe^{0})\right)\\
&\stackrel{P}{\to}{\rm E}\left[\frac{A-e(\bX;\balp^{*})}{e(\bX;\balp^{*})(1-e(\bX;\balp^{*}))}\left(Y-m_{A}(\bX;\bbe^{0})\right)\right]\\
&={\rm E}\left[\frac{A-e(\bX;\balp^{*})}{e(\bX;\balp^{*})(1-e(\bX;\balp^{*}))}\left({\rm E}\left[Y\mid A,\bX\right]-m_{A}(\bX;\bbe^{0})\right)\right]\\
&=0,
\end{align*}
and similarly, 
\begin{align*}
\frac{\partial}{\partial\balp^{\top}}B_n(\balp^{*},\bbe^{0})&=-\frac{1}{n}\sum_{i=1}^n \frac{(A_{i}-e(\bX_{i};\balp^{*}))^2}{(e(\bX_{i};\balp^{*})(1-e(\bX_{i};\balp^{*})))^2}\left(\frac{\partial}{\partial\balp^{\top}}e(\bX_{i};\balp^{*})\right)\left(Y_{i}-m_{A_{i}}(\bX_{i};\bbe^{0})\right)\\
&\stackrel{P}{\to}0
\end{align*}
under some mild conditions. Therefore, (\ref{eql2}) becomes
\begin{align*}
&\int\left|1-\exp\left\{\lambda B_n(\balp,\bbe)\right\}\right|\exp\left\{-n(f_{n}(\balp)+f_{n}(\bbe))\right\}p(\balp)p(\bbe)d\balp d\bbe\\
&<\left\{\sup_{\bbe\in A_{\varepsilon}^{\bbe}}\left|\bbe-\bbe^{0}\right|\right\}\times\left|\left|\frac{\partial}{\partial\bbe^{\top}}B_n(\balp^{*},\bbe^{0})\right|\right|\times(1+\varepsilon)^2+o_{p}(1).
\end{align*}
By taking $\varepsilon\, (>0)$ sufficiently small, under {\bf (C.2)}, the right-hand side becomes arbitrarily close to $0$. 
The same argument holds for $R_4$. 
Therefore, we have 
$$
\int\left|\widetilde{\pi}_{n,\lambda}(\balp,\bbe)-p_{n}(\balp|D)p_{n}(\bbe|D)\right|d\balp d\bbe\stackrel{}{\to}0.
$$
From this result, under $n\to\infty$, it follows that 
$$
\left|z_{n}^{\balp,\bbe}-\int p_n(\balp|D)p_n(\bbe|D)d\balp d\bbe\right|=\left|\int\left\{\widetilde{\pi}_{n,\lambda}(\balp,\bbe)-p_n(\balp|D)p_n(\bbe|D)\right\}d\balp d\bbe\right|
\to0,
$$
which completes the proof.

\subsection{Proof of Theorem 1}\label{secB.2}
When the outcome model is correctly specified ($\bbe^{*}=\bbe^{0}$), from Lemma 1, (\ref{thm1}) obviously holds.

When the propensity score model is correctly specified ($\balp^{*}=\balp^{0}$), considering the following formula:
\begin{align*}
&\left|{\rm E}_{\pi_{n,\lambda}(\bbe)}\left[\frac{1}{n}\sum_{i=1}^{n}\left\{m_{1}(\bX_{i};\bbe)-m_{0}(\bX_{i};\bbe)\right\}\right]-\tau\right| \\
&\leq\left|\hat{\tau}_{IPW}-{\rm E}_{\pi_{n,\lambda}(\bbe)}\left[\frac{1}{n}\sum_{i=1}^{n}\left\{m_{1}(\bX_{i};\bbe)-m_{0}(\bX_{i};\bbe)\right\}\right]\right|+\left|\hat{\tau}_{IPW}-\tau\right|,
\end{align*}
where
$$
\hat{\tau}_{IPW}=\frac{1}{n}\sum_{i=1}^{n}\left\{\frac{A_{i}Y_{i}}{e(\bX_{i};\balp^{0})}-\frac{(1-A_{i})Y_{i}}{1-e(\bX_{i};\balp^{0})}\right\}.
$$
Since the IPW estimator is consistent under some mild conditions, $\hat{\tau}_{IPW}\stackrel{P}{\to}\tau$, we only consider the first term of the above inequality:
\begin{align}
&\left| \frac{1}{n}\sum_{i=1}^{n}\left\{\frac{A_{i}Y_{i}}{e(\bX_{i};\balp^{0})}-\frac{(1-A_{i})Y_{i}}{1-e(\bX_{i};\balp^{0})}\right\}-{\rm E}_{\pi_{n,\lambda}(\bbe)}\left[\frac{1}{n}\sum_{i=1}^{n}\left\{m_{1}(\bX_{i};\bbe)-m_{0}(\bX_{i};\bbe)\right\}\right]\right|\nonumber\\
&=\left| {\rm E}_{\pi_{n,\lambda}(\bbe)}\left[\frac{1}{n}\sum_{i=1}^{n}\left\{\frac{A_{i}Y_{i}}{e(\bX_{i};\balp^{0})}-m_{1}(\bX_{i};\bbe)-\left(\frac{(1-A_{i})Y_{i}}{1-e(\bX_{i};\balp^{0})}-m_{0}(\bX_{i};\bbe)\right)\right\}\right]\right|. \label{eq_a21}
\end{align}

First, considering the first two components of (\ref{eq_a21}) (i.e., for $a=1$):
\begin{align*}
&{\rm E}_{\pi_{n,\lambda}(\bbe)}\left[\frac{1}{n}\sum_{i=1}^{n}\left\{\frac{A_{i}Y_{i}}{e(\bX_{i};\balp^{0})}-m_{1}(\bX_{i};\bbe)\right\}\right]\\
&=\int \frac{1}{n}\sum_{i=1}^{n}\left\{\frac{A_{i}Y_{i}}{e(\bX_{i};\balp^{0})}-m_{1}(\bX_{i};\bbe)\right\}\frac{\exp\left\{\lambda B_n(\balp,\bbe)\right\}p_{n}(d\balp|D)p_{n}(d\bbe|D)}{z_{n}^{\balp,\bbe}}\\
&=\int \frac{1}{n}\sum_{i=1}^{n}\left\{\frac{A_{i}Y_{i}}{e(\bX_{i};\balp^{0})}-\frac{A_{i}m_{1}(\bX_{i};\bbe)}{e(\bX_{i};\balp^{0})}\right\}\frac{\exp\left\{\lambda B_n(\balp,\bbe)\right\}p_{n}(d\balp|D)p_{n}(d\bbe|D)}{z_{n}^{\balp,\bbe}}+o_{p}(1)\\
&=\int \frac{1}{n}\sum_{i=1}^{n}\frac{A_{i}}{e(\bX_{i};\balp^{0})}\left(Y_{i}-m_{A_{i}}(\bX_{i};\bbe)\right)\frac{\exp\left\{\lambda B_n(\balp,\bbe)\right\}p_{n}(d\balp|D)p_{n}(d\bbe|D)}{z_{n}^{\balp,\bbe}}+o_{p}(1)\\
&=\int \frac{1}{n}\sum_{i=1}^{n}\left[\frac{A_{i}}{e(\bX_{i};\balp)}\left(Y_{i}-m_{A_{i}}(\bX_{i};\bbe)\right)-\frac{A_{i}}{e(\bX_{i};\balp)^2}\left(Y_{i}-m_{A_{i}}(\bX_{i};\bbe)\right)\left(\frac{\partial}{\partial\balp^{\top}}e(\bX_{i};\balp)\right)\right.\\
&\hspace{0.5cm}\times\left.
\vphantom{\frac{A_{i}}{e(\bX_{i};\balp)}}   
(\balp-\balp^{0})\right]\frac{\exp\left\{\lambda B_n(\balp,\bbe)\right\}p_{n}(d\balp|D)p_{n}(d\bbe|D)}{z_{n}^{\balp,\bbe}}+o_{p}(1).
\end{align*}
Here, the last equation is a Taylor expansion with respect to $\balp^{0}$ around $\balp$. From the same discussion, the last two components of (\ref{eq_a21}) (i.e., for $a=0$) becomes
\begin{align*}
&{\rm E}_{\pi_{n,\lambda}(\bbe)}\left[\frac{1}{n}\sum_{i=1}^{n}\left\{\frac{(1-A_{i})Y_{i}}{1-e(\bX_{i};\balp^{0})}-m_{0}(\bX_{i};\bbe)\right\}\right]\\
&=\int\frac{1}{n}\sum_{i=1}^{n}\left[\frac{1-A_{i}}{1-e(\bX_{i};\balp)}\left(Y_{i}-m_{A_{i}}(\bX_{i};\bbe)\right)+\frac{1-A_{i}}{(1-e(\bX_{i};\balp))^2}\left(Y_{i}-m_{A_{i}}(\bX_{i};\bbe)\right)\left(\frac{\partial}{\partial\balp^{\top}}e(\bX_{i};\balp)\right)\right.\\
&\hspace{0.5cm}\left.\times
\vphantom{\frac{A_{i}}{e(\bX_{i};\balp)}}   
(\balp-\balp^{0})\right]\frac{\exp\left\{\lambda B_n(\balp,\bbe)\right\}p_{n}(d\balp|D)p_{n}(d\bbe|D)}{z_{n}^{\balp,\bbe}}+o_{p}(1).
\end{align*}
Therefore,
\begin{align}
&{\rm E}_{\pi_{n,\lambda}(\bbe)}\left[\frac{1}{n}\sum_{i=1}^{n}\left\{\frac{A_{i}Y_{i}}{e(\bX_{i};\balp^{0})}-m_{1}(\bX_{i};\bbe)-\left(\frac{(1-A_{i})Y_{i}}{1-e(\bX_{i};\balp^{0})}-m_{0}(\bX_{i};\bbe)\right)\right\}\right]\nonumber\\
&=\int\frac{1}{n}\sum_{i=1}^{n}\left[\frac{A_{i}}{e(\bX_{i};\balp)}\left(Y_{i}-m_{A_{i}}(\bX_{i};\bbe)\right)-\frac{1-A_{i}}{1-e(\bX_{i};\balp)}\left(Y_{i}-m_{A_{i}}(\bX_{i};\bbe)\right)\right.\nonumber\\
&\hspace{0.5cm}-\frac{A_{i}}{e(\bX_{i};\balp)^2}\left(Y_{i}-m_{A_{i}}(\bX_{i};\bbe)\right)\left(\frac{\partial}{\partial\balp^{\top}}e(\bX_{i};\balp)\right)(\balp-\balp^{0})\nonumber\\
&\hspace{0.5cm}\left.-\frac{1-A_{i}}{(1-e(\bX_{i};\balp))^2}\left(Y_{i}-m_{A_{i}}(\bX_{i};\bbe)\right)\left(\frac{\partial}{\partial\balp^{\top}}e(\bX_{i};\balp)\right)(\balp-\balp^{0})\right]\nonumber\\
&\hspace{0.5cm}\times\frac{\exp\left\{\lambda B_n(\balp,\bbe)\right\}p_{n}(d\balp|D)p_{n}(d\bbe|D)}{z_{n}^{\balp,\bbe}}+o_{p}(1)\nonumber\\
&=\int\frac{1}{n}\sum_{i=1}^{n}\left[\left(\frac{A_{i}}{e(\bX_{i};\balp)}-\frac{1-A_{i}}{1-e(\bX_{i};\balp)}\right)\left(Y_{i}-m_{A_{i}}(\bX_{i};\bbe)\right)\right.\nonumber\\
&\hspace{0.5cm}\left.-\left(\frac{A_{i}}{e(\bX_{i};\balp)^2}+\frac{1-A_{i}}{(1-e(\bX_{i};\balp))^2}\right)\left(Y_{i}-m_{A_{i}}(\bX_{i};\bbe)\right)\left(\frac{\partial}{\partial\balp^{\top}}e(\bX_{i};\balp)\right)(\balp-\balp^{0})\right]\nonumber\\
&\hspace{0.5cm}\times\frac{\exp\left\{\lambda B_n(\balp,\bbe)\right\}p_{n}(d\balp|D)p_{n}(d\bbe|D)}{z_{n}^{\balp,\bbe}}+o_{p}(1)\nonumber\\
&=-\int\frac{1}{n}\sum_{i=1}^{n}\left[\left(\frac{A_{i}}{e(\bX_{i};\balp)^2}+\frac{1-A_{i}}{(1-e(\bX_{i};\balp))^2}\right)\left(Y_{i}-m_{A_{i}}(\bX_{i};\bbe)\right)\left(\frac{\partial}{\partial\balp^{\top}}e(\bX_{i};\balp)\right)(\balp-\balp^{0})\right]\nonumber\\
&\hspace{0.5cm}\times\frac{\exp\left\{\lambda B_n(\balp,\bbe)\right\}p_{n}(d\balp|D)p_{n}(d\bbe|D)}{z_{n}^{\balp,\bbe}}+o_{p}(1). \label{ineqa1}
\end{align}
Here, the last equation becomes from the entropic tilting condition (\ref{eq:constraint}).

From the above discussions, (\ref{eq_a21}) becomes
\begin{align*}
&\left| \frac{1}{n}\sum_{i=1}^{n}\left\{\frac{A_{i}Y_{i}}{e(\bX_{i};\balp^{0})}-\frac{(1-A_{i})Y_{i}}{1-e(\bX_{i};\balp^{0})}\right\}-{\rm E}_{\pi_{n,\lambda}(\bbe)}\left[\frac{1}{n}\sum_{i=1}^{n}\left\{m_{1}(\bX_{i};\bbe)-m_{0}(\bX_{i};\bbe)\right\}\right]\right|\\
&\leq\int\frac{1}{n}\sum_{i=1}^{n}\left|\left(\frac{A_{i}}{e(\bX_{i};\balp)^2}+\frac{1-A_{i}}{(1-e(\bX_{i};\balp))^2}\right)\left(Y_{i}-m_{A_{i}}(\bX_{i};\bbe)\right)\left(\frac{\partial}{\partial\balp^{\top}}e(\bX_{i};\balp)\right)\right||\balp-\balp^{0}|\\
&\hspace{0.5cm}\times\frac{\exp\left\{\lambda B_n(\balp,\bbe)\right\}p_{n}(d\balp|D)p_{n}(d\bbe|D)}{z_{n}^{\balp,\bbe}}+o_{p}(1)\\
&\leq\left(\int\left|\left|\frac{1}{n}\sum_{i=1}^{n}\left|\left(\frac{A_{i}}{e(\bX_{i};\balp)^2}+\frac{1-A_{i}}{(1-e(\bX_{i};\balp))^2}\right)\left(Y_{i}-m_{A_{i}}(\bX_{i};\bbe)\right)\left(\frac{\partial}{\partial\balp^{\top}}e(\bX_{i};\balp)\right)\right|\right.\right.\right.\\
&\hspace{0.5cm}\left.
\vphantom{\left|\frac{A_{i}}{e(\bX_{i};\balp)}\right|^2}   
\times \left.\left.\frac{\exp\left\{\lambda B_n(\balp,\bbe)\right\}}{z_{n}^{\balp,\bbe}}\right|\right|^2p_{n}(d\balp|D)p_{n}(d\bbe|D)\right)^{1/2}\left(E_{p_{n}(\balp\mid D)}\left[||\balp-\balp^{0}||^2\right]\right)^{1/2}+o_{p}(1)\\
&\to0,
\end{align*}
where the second inequality is from the H\"{o}lder inequality, and the last convergence is from {\bf (C.3)} and Lemma A.1. Therefore, under sufficient large $n$, (\ref{thm1}) holds.

\section{Benefits of incorporating entropic tilting} \label{app:npb}
From the same discussion as in Section \ref{secB.2}, without using ET (i.e., only using regression model such as BART), the difference between the IPW estimator becomes
\begin{align}
&\int\frac{1}{n}\sum_{i=1}^{n}\left[\left(\frac{A_{i}}{e(\bX_{i};\balp)}-\frac{1-A_{i}}{1-e(\bX_{i};\balp)}\right)\left(Y_{i}-m_{A_{i}}(\bX_{i};\bbe)\right)\right.\nonumber\\
&\hspace{0.5cm}\left.-\left(\frac{A_{i}}{e(\bX_{i};\balp)^2}+\frac{1-A_{i}}{(1-e(\bX_{i};\balp))^2}\right)\left(Y_{i}-m_{A_{i}}(\bX_{i};\bbe)\right)\left(\frac{\partial}{\partial\balp^{\top}}e(\bX_{i};\balp)\right)(\balp-\balp^{0})\right]\nonumber\\
&\hspace{0.5cm}\times \frac{p_{n}(d\balp|D)p_{n}(d\bbe|D)}{z_{n}^{\balp}z_{n}^{\bbe}}+o_{p}(1).\label{eqc1}
\end{align}
Considering the Taylor expansion of the first term:
\begin{align*}
&\frac{A_{i}}{e(\bX_{i};\balp)}-\frac{1-A_{i}}{1-e(\bX_{i};\balp)}\\
&=\frac{A_{i}}{e(\bX_{i};\balp^{0})}-\frac{1-A_{i}}{1-e(\bX_{i};\balp^{0})}-\left(\frac{A_{i}}{e(\bX_{i};\balp^{0})^2}+\frac{1-A_{i}}{(1-e(\bX_{i};\balp^{0}))^2}\right)\left(\frac{\partial}{\partial\balp^{\top}}e(\bX_{i};\balp^{0})\right)(\balp-\balp^{0}).
\end{align*}
Therefore, (\ref{eqc1}) becomes
\begin{align}
&\int\frac{1}{n}\sum_{i=1}^{n}\left[\left(\frac{A_{i}}{e(\bX_{i};\balp^{0})}-\frac{1-A_{i}}{1-e(\bX_{i};\balp^{0})}\right)\left(Y_{i}-m_{A_{i}}(\bX_{i};\bbe)\right)\right.\nonumber\\
&\hspace{0.5cm}\left.-\left\{\left(\frac{A_{i}}{e(\bX_{i};\balp^{0})^2}+\frac{1-A_{i}}{(1-e(\bX_{i};\balp^{0}))^2}\right)\left(Y_{i}-m_{A_{i}}(\bX_{i};\bbe)\right)\left(\frac{\partial}{\partial\balp^{\top}}e(\bX_{i};\balp^{0})\right)\right.\right.\nonumber\\
&\hspace{0.5cm}\left.\left.+\left(\frac{A_{i}}{e(\bX_{i};\balp)^2}+\frac{1-A_{i}}{(1-e(\bX_{i};\balp))^2}\right)\left(Y_{i}-m_{A_{i}}(\bX_{i};\bbe)\right)\left(\frac{\partial}{\partial\balp^{\top}}e(\bX_{i};\balp)\right)\right\}(\balp-\balp^{0})\right]\nonumber\\
&\hspace{0.5cm}\times \frac{p_{n}(d\balp|D)p_{n}(d\bbe|D)}{z_{n}^{\balp}z_{n}^{\bbe}}+o_{p}(1)\nonumber\\
&=\int\frac{1}{n}\sum_{i=1}^{n}\left[\left(\frac{A_{i}}{e(\bX_{i};\balp^{0})}-\frac{1-A_{i}}{1-e(\bX_{i};\balp^{0})}\right)\left(Y_{i}-m_{A_{i}}(\bX_{i};\bbe^{0})\right)\right.\nonumber\\
&\hspace{0.5cm}-\left(\frac{A_{i}}{e(\bX_{i};\balp^{0})}-\frac{1-A_{i}}{1-e(\bX_{i};\balp^{0})}\right)\left(m_{A_{i}}(\bX_{i};\bbe)-m_{A_{i}}(\bX_{i};\bbe^{0})\right)\nonumber\\
&\hspace{0.5cm}\left.-\left\{\left(\frac{A_{i}}{e(\bX_{i};\balp^{0})^2}+\frac{1-A_{i}}{(1-e(\bX_{i};\balp^{0}))^2}\right)\left(Y_{i}-m_{A_{i}}(\bX_{i};\bbe)\right)\left(\frac{\partial}{\partial\balp^{\top}}e(\bX_{i};\balp^{0})\right)\right.\right.\nonumber\\
&\hspace{0.5cm}\left.\left.+\left(\frac{A_{i}}{e(\bX_{i};\balp)^2}+\frac{1-A_{i}}{(1-e(\bX_{i};\balp))^2}\right)\left(Y_{i}-m_{A_{i}}(\bX_{i};\bbe)\right)\left(\frac{\partial}{\partial\balp^{\top}}e(\bX_{i};\balp)\right)\right\}(\balp-\balp^{0})\right]\nonumber\\
&\hspace{0.5cm}\times \frac{p_{n}(d\balp|D)p_{n}(d\bbe|D)}{z_{n}^{\balp}z_{n}^{\bbe}}+o_{p}(1).\label{eqc2}
\end{align}

Whereas, with ET and under correct specification of the outcome model, (\ref{ineqa1}) becomes
\begin{align}
&-\int\frac{1}{n}\sum_{i=1}^{n}\left[\left(\frac{A_{i}}{e(\bX_{i};\balp)^2}+\frac{1-A_{i}}{(1-e(\bX_{i};\balp))^2}\right)\left(Y_{i}-m_{A_{i}}(\bX_{i};\bbe)\right)\left(\frac{\partial}{\partial\balp^{\top}}e(\bX_{i};\balp)\right)(\balp-\balp^{0})\right]\nonumber\\
&\hspace{0.5cm}\times\frac{\exp\left\{\lambda B_n(\balp,\bbe)\right\}p_{n}(d\balp|D)p_{n}(d\bbe|D)}{z_{n}^{\balp,\bbe}}+o_{p}(1)\nonumber\\
&=-\int\frac{1}{n}\sum_{i=1}^{n}\left[\left(\frac{A_{i}}{e(\bX_{i};\balp)^2}+\frac{1-A_{i}}{(1-e(\bX_{i};\balp))^2}\right)\left(Y_{i}-m_{A_{i}}(\bX_{i};\bbe)\right)\left(\frac{\partial}{\partial\balp^{\top}}e(\bX_{i};\balp)\right)(\balp-\balp^{0})\right]\nonumber\\
&\hspace{0.5cm}\times \frac{p_{n}(d\balp|D)p_{n}(d\bbe|D)}{z_{n}^{\balp} z_{n}^{\bbe}}+o_{p}(1)\label{eqc3}
\end{align}
asymptotically (see Lemma 1).

From (\ref{eqc2}) and (\ref{eqc3}), the difference in convergence rates between the cases with and without ET is derived from the second term of (\ref{eqc2}):
\begin{align*}
&\int\frac{1}{n}\sum_{i=1}^{n}\left(\frac{A_{i}}{e(\bX_{i};\balp^{0})}-\frac{1-A_{i}}{1-e(\bX_{i};\balp^{0})}\right)\left(m_{A_{i}}(\bX_{i};\bbe)-m_{A_{i}}(\bX_{i};\bbe^{0})\right)\frac{p_{n}(d\bbe|D)}{z_{n}^{\bbe}}\\
&=\int\frac{1}{n}\sum_{i=1}^{n}\left(\frac{A_{i}-e(\bX_{i};\balp^{0})}{e(\bX_{i};\balp^{0})(1-e(\bX_{i};\balp^{0}))}\right)\left(m_{A_{i}}(\bX_{i};\bbe)-m_{A_{i}}(\bX_{i};\bbe^{0})\right)\frac{p_{n}(d\bbe|D)}{z_{n}^{\bbe}}.
\end{align*}
Following the discussion in \cite{dukes2024doubly}, the convergence rate of this term may be slower than that of the other terms because:
\begin{enumerate}
\item $n^{-1}\sum_{i=1}^{n}\left(\frac{A_{i}-e(\bX_{i};\balp^{0})}{e(\bX_{i};\balp^{0})(1-e(\bX_{i};\balp^{0}))}\right)\left(Y_{i}-m_{A_{i}}(\bX_{i};\bbe^{0})\right)=O_{p}(1/\sqrt{n})$, and
\item the third terms of (\ref{eqc2}) and (\ref{eqc3}) contain cross terms involving $\left(Y_{i}-m_{A_{i}}(\bX_{i};\bbe)\right)\times(\balp-\balp^{0})$ which converge faster than $O_{p}(1/\sqrt{n})$.
\end{enumerate}
Therefore, without using ET, the convergence rate may be primarily determined by the outcome model. For instance, under BART, the convergence rate is slower than $\sqrt{n}$--order \citep{rovckova2019theory}.

Thus, it is expected that the proposed Bayesian DR method is clearly advantageous in terms of convergence rate reduction when the propensity score model is correctly specified. Additionally, from Lemma 1, it is expected that the posterior distribution for the outcome model becomes the same both with and without using ET asymptotically, even if the propensity score model is misspecified. These points are also confirmed by the simulation experiments presented in the main manuscript.

\section{Additional simulation experiments}
\subsection{High-dimensional setting}\label{add_sim_HD}
In this section, we consider a high-dimensional setting for covariates. To address this, we apply our proposed method using a shrinkage prior, such as the horseshoe prior \citep{makalic2015simple}. Note that Saarela’s method cannot accommodate such a shrinkage prior due to its estimation procedure.

In the data-generating mechanism, we add 40 irrelevant covariates that are unrelated to both the propensity score and the outcome. Specifically, these covariates are generated as $X_{ji}\sim N(u_{j},1)$, $u_{j}\sim Unif(-1,1)$ ($j=1,\dots,40$).

We only show an one-shot result when $n=200$. When both the propensity score and outcome models are correctly specified, the posterior mean and standard deviation of Saarela's method are 116.52 (90.31). In contrast, the G-formula and our proposed method yield 110.07 (0.46) and 110.06 (0.45), respectively. Both two methods clearly mitigate the impact of high dimensionality through the use of shrinkage priors.

\subsection{\citeauthor{Ka2007}'s misspecification setting}\label{add_sim_KS}
In this section, we additionally consider the situation in which the outcome model is misspecified following the misspecification manner in \cite{Ka2007}.
Specifically, we use the following misspecified covariates for the outcome model:
$$
\left(\exp\left\{\frac{X_{1}}{2}\right\}, 10+\frac{X_{2}}{1+\exp\left\{X_{1}\right\}},\left(0.6+\frac{X_{1}X_{3}}{25}\right)^3,\left(20+X_{1}+X_{4}\right)^2\right)
$$
Here, we use the covariates after standardization.
The other settings are the same as those in the simulation experiments in the main manuscript.
As is obvious, this misspecification is severe and extreme compared with that in the main manuscript \citep{robins2007comment}.

The results are summarized in Table \ref{tab3}; \citeauthor{Bang2005doubly}'s DR estimator and the G-formula show similar tendencies in the main manuscript.
Our proposed methods decrease bias compared with the G-formula; however, the CP exhibits undercoverage.
The pruning algorithm can further improve bias, RMSE, and CP; however, it still exhibits undercoverage.
The main results and this additional simulation experiment show that our proposed methodology can improve bias, at least compared with the G-formula; however, CP depends on the misspecification pattern of the outcome model.

\begin{table}[h]
\begin{center}
\caption{Summary of causal effect estimates under outcome model misspecification:\ The number of iteration is $2000$ and the true ATE is 110. The absolute bias (ABias), empirical standard error (ESE), root mean squared error (RMSE), coverage probability (CP), and average length of credible interval (AvL) of the estimated causal effects across 2000 iterations are summarized by estimation method (``Method" column).}
\label{tab3}
\begin{tabular}{c|ccccc}\hline
{\bf Method}&\multicolumn{5}{|c}{\bf Sample size:\ $n=500$} \\ \cline{2-6}
& ABias & ESE & RMSE & CP & AvL \\ \hline
DR        & 0.019 & 1.266 & 1.266 & 100 & 7.103  \\ \hline
G-formula & 3.439 & 1.494 & 3.750 & 37.7 & 5.927  \\ \hline
Proposed  & 1.565 & 1.092 & 1.909 & 88.8 & 5.879 \\ \hline\hline
\begin{tabular}{c}{Proposed} \\ {(pruning)} \end{tabular} & 0.327 & 1.333 & 1.372 & 91.3 & 4.640 \\ \hline
\end{tabular}
\end{center}
{\footnotesize{
DR:\ Ordinaly non-Bayesian DR estimator that is asymptotically equivalent to \cite{Bang2005doubly}; 
G-formula:\ Bayesian G-formula based method discussed in \cite{daniels2023bayesian}; 
Pruning:\ Using sample pruning algorithm for our proposed method described in Section \ref{sec5.3}.1\\
For our proposed method, sequential Monte Carlo with replacement is conducted. For all Bayesian methods, 20,000 posterior draws are used.
}}
\end{table}

\section{Importance weighting with a single sensitivity parameter}
\label{app:impwgt_kuan}
When $\xi_{i}\equiv\xi$, we can consider another option for the sampling algorithm for $\xi$.
Let $f(Y_i\mid A_i,\bX_i;\bbe,\xi)$ be the corresponding negative log-likelihood function of the outcome model including $\xi$. 
Then, the joint posterior of $(\balp, \bbe)$ is proportional to 
\begin{align}
&\pi(\balp)\pi(\bbe)\prod_{i=1}^n \exp\left\{- f(A_i\mid \bX_i;\balp) \right\} \int \exp\left\{-f(Y_i\mid A_i,\bX_i;\bbe,\xi) \right\}g(\xi)d\xi \nonumber \\
&=\pi(\balp)\pi(\bbe)\exp\left\{-\sum_{i=1}^{n}f(A_i\mid \bX_i;\balp) \right\} \int \exp\left\{-\sum_{i=1}^{n}f(Y_i\mid A_i,\bX_i;\bbe,\xi) \right\}g(\xi)d\xi \label{eq:pos-sensitivity2},
\end{align}
where $g(\xi)$ is a density function of $\xi$.
In the same manner as in the main manuscript, the posterior samples from the above posterior can be generated by an importance sampling with important weight proportional to 
$$
\frac{\int \exp\left\{-\sum_{i=1}^{n}f(Y_i\mid A_i,\bX_i;\bbe,\xi)\right\}g(\xi)d\xi}{\exp\left\{-\sum_{i=1}^{n}f(Y_i\mid A_i,\bX_i;\bbe)\right\}}.
$$
Assuming that random sample generation from $g(\cdot)$ is tractable, we propose the following procedure to obtain the posterior sample from (\ref{eq:pos-sensitivity2}).

\begin{aalgo} (Importance sampling approach for sensitivity analysis)
For $s=1,\ldots,S$, generate $M$ samples from $g(\cdot)$, denoted by $\xi^{(m)}$ for $m=1,\ldots,M$ and compute the weight as 
$$
w^{(s)}\equiv \frac{M^{-1}\sum_{m=1}^M\exp\left\{-\sum_{i=1}^{n}f(Y_i\mid A_i,\bX_i;\bbe^{(s)},\xi^{(m)})\right\}}{\exp\left\{-\sum_{i=1}^{n}f(Y_i\mid A_i,\bX_i;\bbe^{(s)})\right\}}.
$$
Then, the posterior samples can be approximated by the particles $(\balp^{(s)},\bbe^{(s)})$ and its weight $w^{(s)}$ for $s=1,\ldots,S$.
\end{aalgo}

\section{Entropic tilting using propensity score subclassification} \label{app:subclass}
Propensity score subclassification is known as one of the confounder adjustment methods using the propensity score. As mentioned in \cite{imbens2015causal}, propensity score subclassification is more stable than the IPW estimator because extreme weights can be smoothed within each stratum.

Using the estimated propensity score $\hat{e}_{i}\equiv e(\bX_{i};\hat{\bld{\balp}})$, the subclassification estimator can be represented as:
\begin{align}
\label{Sub_est}\frac{1}{n}\sum_{i=1}^{n}\sum_{k=1}^{K}\left(\frac{A_{i}}{n_{k1}/n_{k+}}-\frac{(1-A_{i})}{1-n_{k1}/n_{k+}}\right)Y_{i}{\rm I}_{\{\hat{c}_{k-1}\leq \hat{e}_{i}<\hat{c}_{k}\}},
\end{align}
where $K$ is the number of strata, and each stratum is constructed as
$(\hat{c}_{0},\, \hat{c}_{1})\cup\bigcup_{k=2}^{K}[\hat{c}_{k-1},\, \hat{c}_{k})=(0,\, 1)$, with $0=\hat{c}_{0}<\hat{c}_{1}<\cdots<\hat{c}_{K}=1$. Here, $n_{k+}$ as the sample size within the interval $[\hat{c}_{k-1}, \hat{c}_{k})$, and $n_{1k}$ and $n_{0k}$ as the sample sizes for $A=1$ and $A=0$ within this interval, respectively (i.e., $n_{k+} = n_{1k} + n_{0k}$). Typically, strata are constructed as equal-frequency strata \citep{orihara2021determination}, where $n_{+} \equiv n_{k+} = n / K$. Hereafter, we note ${\rm I}_{\{\hat{c}_{k-1}\leq \hat{e}_{i}<\hat{c}_{k}\}}={\rm I}_{k}(\bX_{i})$ and the number of strata $K$ does not depend on the sample size $n$.

Compared with (\ref{ipw1}), $n_{k1}/n_{+}$ in (\ref{Sub_est}) can be viewed as the propensity score for each stratum. Therefore, modifying the ET condition (\ref{eq:balancing}):
\begin{align*}
B_n(\balp,\bbe) &\equiv \frac{1}{n} \sum_{i=1}^{n} \frac{A_{i} - e(\bX_{i}; \balp)}{e(\bX_{i}; \balp)(1 - e(\bX_{i}; \balp))} \left(Y_{i} - m_{A_{i}}(\bX_{i}; \bbe)\right) \\
&= \frac{1}{n} \sum_{i=1}^{n} \left(\frac{A_{i}}{e(\bX_{i}; \balp)} - \frac{1 - A_{i}}{1 - e(\bX_{i}; \balp)}\right) \left(Y_{i} - m_{A_{i}}(\bX_{i}; \bbe)\right),
\end{align*}
the ET condition based on propensity score subclassification becomes:
\begin{align}
\label{ET_sub}
B_n^{Sub}(\balp, \bbe) = \frac{1}{n} \sum_{i=1}^{n} \sum_{k=1}^{K} \left(\frac{A_{i}}{n_{k1}/n_{+}} - \frac{1 - A_{i}}{1 - n_{k1}/n_{+}}\right) \left(Y_{i} - m_{A_{i}}(\bX_{i}; \bbe)\right) {\rm I}_{k}(\bX_{i}).
\end{align}
In fact, using (\ref{ET_sub}), the proposed Bayesian procedure also achieves (approximately) double robustness, based on the same discussion presented in the main manuscript.

As mentioned in the main manuscript, our proposed method can explicitly describe a posterior distribution. Therefore, as discussed in \cite{orihara2024bayesian}, an algorithm for guessing the number of strata using reversible jump MCMC can be applied.

\end{document}